\documentclass[authoryear,11pt,3p]{elsarticle}

\usepackage{tabu}
\usepackage{caption}
\usepackage{prettyref}

\usepackage{graphicx}

\usepackage{xcolor}

\usepackage{amssymb}

\usepackage{amsfonts}
\usepackage{rotating}
\usepackage{multirow}
\usepackage{epstopdf}
\usepackage{float}

\usepackage{apalike}
\usepackage{xcolor}

\usepackage{amssymb,amsmath}

\newcommand{\gen}{$RrINAR(\mathcal{M,A,P})$}
\newcommand{\geng}{$RrNGINAR(\mathcal{M,A,P})$}
\newcommand{\gengo}{$RrNGINAR_1(\mathcal{M,A,P})$}
\newcommand{\gengm}{$RrNGINAR_{max}(\mathcal{M,A,P})$}
\newcommand{\gengoTwo}{$R2NGINAR_1(\mathcal{M,A,P})$}
\newcommand{\gengmTwo}{$R2NGINAR_{max}(\mathcal{M,A,P})$}
\newcommand{\gengoThree}{$R3NGINAR_1(\mathcal{M,A,P})$}
\newcommand{\gengmThree}{$R3NGINAR_{max}(\mathcal{M,A,P})$}
\newcommand{\inar}{$INAR$}
\newcommand{\envst}{\emph{RENES}}

\makeatletter
\def\ps@pprintTitle{%
  \let\@oddhead\@empty
  \let\@evenhead\@empty
  \let\@oddfoot\@empty
  \let\@evenfoot\@oddfoot
}
\makeatother

\begin{document}

\newtheorem{dfn}{Definition}
\newtheorem{rem}{Remark}
\newtheorem{thm}{Theorem}
\newtheorem{lma}[thm]{Lemma}
\newtheorem{crl}[thm]{Corollary}
\newdefinition{rmk}{Remark}
\newdefinition{example}{Example}
\newproof{pf}{Proof}
\newproof{pot}{Proof of Theorem}
\def\ack{\section*{Acknowledgements}%
  \addtocontents{toc}{\protect\vspace{6pt}}%
  \addcontentsline{toc}{section}{Acknowledgements}%
}

\begin{frontmatter}



\title{On Generalized Random Environment \inar\ Models of Higher Order: Estimation of Random Environment States}



\author{Bogdan A. Pirkovi\' c}
\ead{apirkovic@yahoo.com }
{{\address{University of Kragujevac, Faculty of Science, Department of Mathematics and Informatics, Radoja Domanovi\' ca 12, 34000 Kragujevac, Serbia}}}

\author{Petra N. Laketa}
\ead{laketa@karlin.mff.cuni.cz}
\address{Charles University, Faculty of Mathematics and Physics, Ke Karlovu 3, 121 16 Praha 2, Czech republic}

\author{Aleksandar S. Nasti\'c}
\ead{anastic78@gmail.com}
\address{University of Ni\v s, Faculty of Sciences and Mathematics, Vi\v segradska 33, 18000 Ni\v s, Serbia}


\begin{abstract}
The behavior of a generalized random environment integer-valued autoregressive model of higher order with geometric marginal distribution {and negative binomial thinning operator} (abbrev. \geng) is dictated by a realization $\{z_n\}_{n=1}^\infty$ of an auxiliary Markov chain called random environment process. Element $z_n$ represents a state of the environment in moment $n\in\mathbb{N}$ and determines three different parameters of the model in that moment. In order to use \geng\ model, one first needs to estimate $\{z_n\}_{n=1}^\infty$, which was so far done by K-means data clustering. We argue that this approach ignores some information and performs poorly in certain situations. We propose a new method for estimating $\{z_n\}_{n=1}^\infty$, which includes the data transformation preceding the clustering, in order to reduce the information loss. To confirm its efficiency, we compare this new approach with the usual one when applied on the simulated and the real-life data, and notice all the benefits obtained from our method.
\end{abstract}

\begin{keyword}
\geng, Random environment, K-means, $INAR$, estimation

\MSC 62M10

\end{keyword}

\end{frontmatter}

\section{Introduction}

Integer-valued autoregressive (\inar) models appeared for the first time in \cite{McKenzie} and \cite{AlAlzaid}. Over the time, they showed to be a very useful tool for describing the integer-valued data. One may, for example, apply \inar\ models to describe the monthly number of rainy days, crime cases, newborn individuals of one species. It becomes clear that such data may be found in any area. For that reason, numerous \inar\ models have been proposed and studied in the literature. \inar\ models are based on so called \emph{thinning operator}, which to a given integer-valued random variable $X$ assigns the sum of $X$ independent identically distributed random variables. The distribution of the auxiliary random variables determines the type of the thinning operator. Some of the models with different thinning operators may be found in \cite{AB1994}, \cite{Latour1998}, \cite{ZengBasawaData2006,ZengBasawaData2007} and {\cite{RBN2009}}. Another variety of \inar\ models arise from considering different marginal distributions, see for example \cite{McKenzie1986}, \cite{Al-OshAly1992}, \cite{AA1993} and \cite{BR2010}.\\

We focus on the recent random environment \inar\ models that appeared for the first time in \cite{NLR2016} and are flexible towards the environment conditions changes. The behavior of these models is ruled by a Markov chain $\{Z_n\}_{n=1}^\infty$, called \emph{random environment process}. The elements of the random environment process are also called \emph{(random) environment states}. To apply a random environment \inar\ model, one must first estimate the environment states. In  \cite{NLR2016} this was done using clustering methods, in particular K-means introduced by \cite{HW1979}. However, using K-means for this purpose induces a certain loss of information, as we discuss below, and may lead to poor performance of the model. In order to estimate environment states as accurate as possible, we propose a new random environment estimation (abbrev. \envst) method. To avoid the confusion, it is important to emphasize that we do not introduce a new method for estimating the parameters of random environment \inar\ models, but only a new method for estimating $\{z_n\}_{n=1}^\infty$. However, the estimators of random environment \inar\ models parameters
 are defined under the assumption that $\{z_n\}_{n=1}^\infty$ is known in advance, meaning that a different approach for estimating $\{z_n\}_{n=1}^\infty$ will for sure imply the difference in the parameter estimates.\\

\begin{figure}[!tbp]
  \centering
    \includegraphics[width=0.45\textwidth, height=4.2cm]{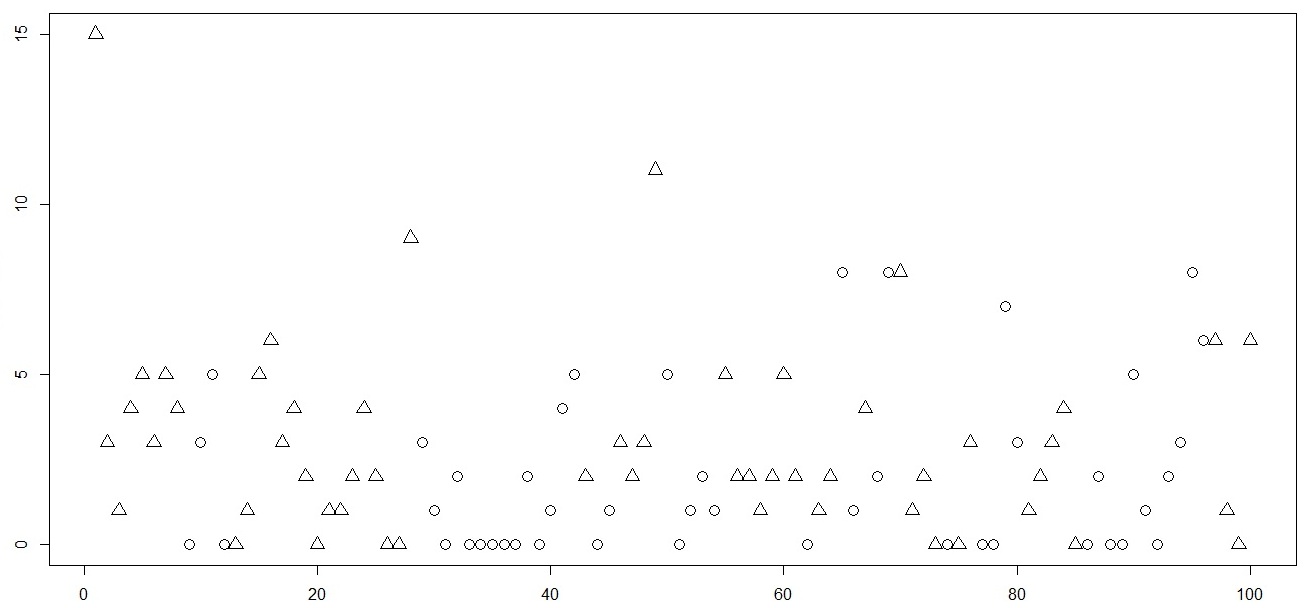}
    \includegraphics[width=0.45\textwidth, height=4.2cm]{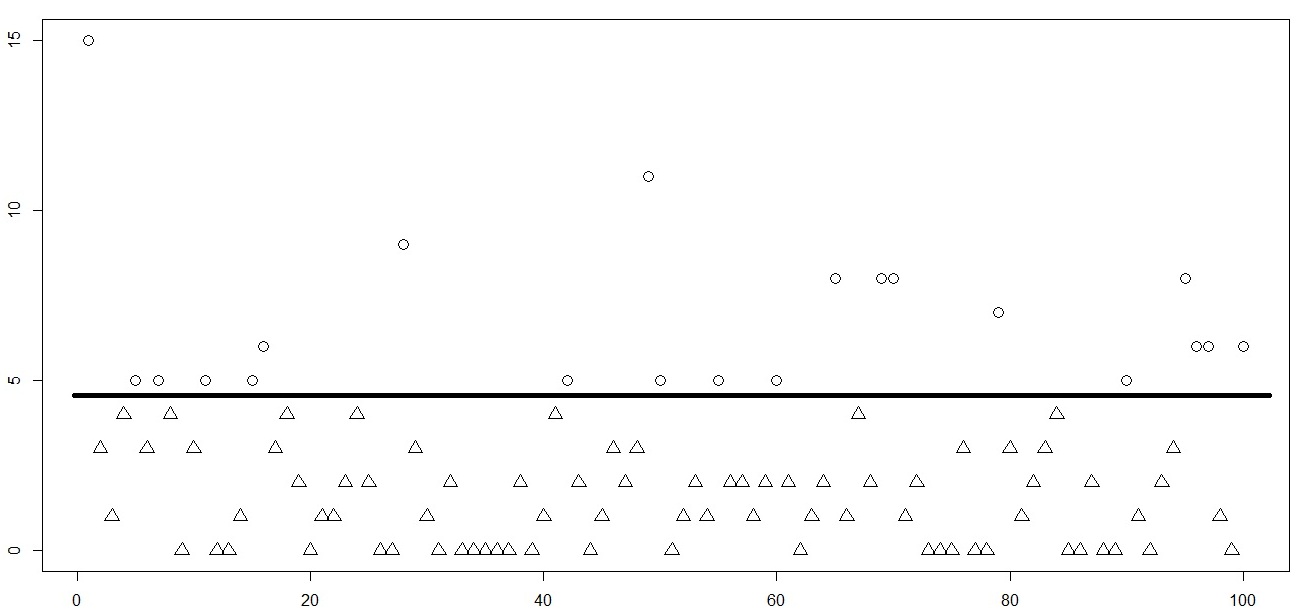}
    \caption{{Simulated $R2NGINAR(1)$ time series: different states presented with a different symbol --- a circle or a triangle. On the left-hand side we see the exact states of the simulated data, while on the right-hand there are the estimated states, obtained by the application of standard K-means method to the simulated data. The estimated states are obviously divided by a horizontal strip, unlike the exact states.}}
    \label{prva}
\end{figure}

We need to provide more details on the random environment \inar\ models. We begin with the first order random environment \inar\ model with geometric marginals ($RrNGINAR(1)$) introduced in \cite{NLR2016}. The marginal distribution of the $RrNGINAR(1)$ time series in moment $n$ is determined by the realization of the random environment process $z_n$ recorded in the same moment --- for this reason we write $X_n(z_n)$. Moreover, the distribution of $X_n(z_n)$ is geometric with expectation $\mu_{z_n}\in\{\mu_1,\mu_2,\dots\mu_r\}$. The recursive relation that defines $RrNGINAR(1)$ model is given by
\begin{equation}\label{first order}
X_n(z_n)=\alpha \ast X_{n-1}(z_{n-1})+\varepsilon_n(z_n,z_{n-1}),
\end{equation}
where, $\alpha\ast:X\mapsto\sum_{i=1}^XU_i$ denotes the \emph{negative binomial thinning operator}, that to each integer-valued random variable $X$ assigns the sum of $X$ independent random variables having geometric distribution with mean value $\alpha$. In order to measure the goodness of fit of such defined model, corresponding environment states $z_n$ for all observations must be estimated. This is where the K-means clustering method took place. The predefined number of clusters {was chosen to be} the number of environment states $r$ registered in the observed phenomenon. Each cluster {was} assigned to one state and each sample element {was} assigned to a state depending on the cluster it felt into. In particular, if two different process elements $X_n(z_n)$ and $X_m(z_m)$ belong to the same cluster, then {it was assumed that} $z_n=z_m$. In that way, the sequence of random environment states {was} fully determined. However, this approach shows a serious shortcomings, which are consequence of the fact that only data point value {was} taken into account in K-means clustering method. Once the K-means is performed, graphed representation of the database will be divided by horizontal lines into strips, as shown {in the right-hand panel of Figure~\ref{prva}}. Each strip corresponds to one cluster. This entails that all high values in the database must be located in the same cluster. Similar to this, all low values must be located in the same cluster. {This is, however, not the case with the data simulated from $RrNGINAR(1)$ process. The left-hand panel of Figure~\ref{prva} shows the data simulated from $R2NGINAR(1)$ model. As we can see, it is possible that high data values appear also in the environment conditions different than those assumed for the high data values, so the data is no longer divided by a horizontal strip. K-means totally rules out this possibility.}\\

As an example, consider the number of the new COVID-19 cases per day. As it is known, the weather conditions significantly affected the spread rate, so we {use $RrNGINAR(1)$ model with $r=2$ different environment states: summer and winter.} However, there are some other circumstances undetected or not measurable, that can affect the number of new cases per day, for example public demonstrations, unallowed gatherings of people during vacations or emergence of the new virus strain. In these situations, it would be ideal for clustering method to recognize specific circumstances and keep high values (detected in summertime) in 'summer' cluster. However, standard K-means is incapable to do so. Observing only numerical value of the process realization, K-means might recognize high summertime values as winter occa\-sions, and locate those realizations in the wrong cluster. The same holds for all K-means adaptations familiar so far. Obviously, an improved random environment estimation method is needed in order to solve such problem.\\

In years that followed, few more sophisticated \inar\ models appeared. \cite{NLR2017} defined random environment \inar\ models of higher order. Beside the marginal distribution parameter, authors assumed here that the order of the model is also determined by the environment state in particular moment. Another step ahead was made by \cite{LNR2018}. Beside all the assumptions mentioned above, authors additionally assumed that the thinning parameter value $\alpha_{z_n}$ in moment $n$ depends on the environment state $z_n$ in the same moment.  According to \cite{LNR2018}, $\{X_n(z_n)\}_{n=1}^\infty$ is called a generalized random environment \inar\ model of higher order with geometric marginals and negative binomial thinning operator ($RrNGINAR(\mathcal{M,A,P})$) if its element $X_n(z_n)$ at moment $n\in\mathbb{N}$ is determined by the recursive relation
\begin{equation}\label{model definition}
X_n(z_n)=\begin{cases}
\alpha_{z_n}\ast X_{n-1}(z_{n-1})+\varepsilon_n(z_n,z_{n-1})& \text{w.p.} \ \phi_{1,P_n}^{z_n},\\
\alpha_{z_n}\ast X_{n-2}(z_{n-2})+\varepsilon_n(z_n,z_{n-2})& \text{w.p.} \ \phi_{2,P_n}^{z_n},\\
\vdots\\
\alpha_{z_n}\ast X_{n-P_n}(z_{n-P_n})+\varepsilon_n(z_n,z_{n-P_n})& \text{w.p.} \ \phi_{P_n,P_n}^{z_n},
\end{cases}
\end{equation}
Sets $\mathcal{M}=\{\mu_1,\dots ,\mu_r\}$, $\mathcal{A}=\{\alpha_1,\dots ,\alpha_r\}$, $\mathcal{P}=\{p_1,\dots ,p_r\}$ contain model parameter values --- $\mu_{z_n}$ is the mean of the marginal geometric distribution of $X_n(z_n)$, $\alpha_{z_n}$ is the thinning parameter value and $p_{z_n}$ represents the maximal value that order $P_n$ may take for a fixed state $z_n\in\{1,\dots ,r\}$. There are two different \geng\ models, depending on the way sequence $\{P_n\}_{n=1}^\infty$ is defined. One of them, \gengm\ is constructed so that for each $n\in\mathbb{N}$ it holds $P_n=\max\{p^*_n,p_{z_n}\}$, while for the other one, \gengo\ we have $P_n=1$ if $p^*_n<p_{z_n}$ and $P_n=p_{z_n}$ otherwise. Here $p^*_n=\max\{i\geq 1\colon z_{n-i}=\dots=z_{n-1}\}$ represents the number of predecessors of $z_n$ that are mutually equal.\\

Beside {the} shortcomings mentioned above, we detected another difficulty {when} applying K-means to a generalized random environment $INAR$ process of higher order, as it was done in \cite{LNR2018}. To explain the difficulty, {consider} the simplest case with $r=2$ environment states and suppose similarity between mean values within states, $\mu_1\approx\mu_2$. In other words, the observations inside states are not that much different and are accumulated around parallel horizontal lines that are close to each other. In this situation, it is reasonable to expect the existence of a strip in which points from both environment states will be mixed. The border between states won't be a straight line, but a wavy and jagged line. Taking into account the fact that K-means method separates clusters by straight horizontal lines, it becomes obvious that some improvements are necessary.\\

In this paper we introduce a new \envst\ method for estimation of $\{z_n\}_{n=1}^\infty$, that will eliminate disadvantages mentioned above. The idea of \envst\ method is to transform the data sample that corresponds to the generalized random environment $INAR$ model of higher order before applying clustering. As previously mentioned, all the parameter values $\mu_{z_n}$, $\alpha_{z_n}$ and $P_n$ carry the information about $z_n$. To prevent the information loss, the main goal is to form a three-dimensional sequence, based on real-life data realizations, that mimics the behavior of $\{(\mu_{z_n},\alpha_{z_n},P_n)\}_{n=1}^\infty$. Finally, the K-means algorithm will be applied on such obtained three-dimensional data sequence. It is possible to apply \envst\ method to any other generalized random environment $INAR$ model of higher order (to those that have different marginal distribution and thinning operator), but we focus on \geng\ models, as it was the case in \cite{LNR2018}. In order to confirm the efficiency of \envst , corresponding simulated data sequences are created. Observing the simulations, we can examine whether changes in the number of states and parameter values affect the efficiency of \envst\ method. \\

The structure of this paper is as follows. In Section $2$, a construction of the new random environment estimation (\envst) method, which overcomes problems mentioned above, is presented. Section $3$ provides description of simulations and their properties. Cases with $2$ and $3$ different environment states are described. An extensive simulation study for newly proposed  \envst\ method is given in Section $4$. Results of applying the \envst\ method to the real-life data are given in Section $5$.

\section{Construction of the new \envst\ method}

Consider a given sample $\{X_n\}_{n=1}^N=\{X_n(z_n)\}_{n=1}^N$ of size $N\in\mathbb{N}$ from \geng\ model. In order to construct the method, the main idea is to determine certain kind of pre-estimators $\{\tilde{\mu}_n\}_{n=1}^N$, $\{\tilde{\alpha}_n\}_{n=1}^N$ and $\{\tilde{P}_{n}\}_{n=1}^N$ of parameter sequences $\{\mu_{z_n}\}_{n=1}^N$, $\{\alpha_{z_n}\}_{n=1}^N$ and $\{P_n\}_{n=1}^N$ based only on the realized sample, without knowing the random environment sequence $\{z_n\}_{n=1}^N$. Such obtained three-dimensional sequence $\{(\tilde{\mu}_n,\tilde{\alpha}_n,\tilde{P}_n)\}_{n=1}^N$ is supposed to mimic the behavior of the model parameters over time. Then, clustering the three-dimensional data $\{(\tilde{\mu}_n,\tilde{\alpha}_n,\tilde{P}_n)\}_{n=1}^N$ would produce better esti\-mation of $\{z_n\}_{n=1}^N$ than clustering of the starting sequence $\{x_n\}_{n=1}^N$, since the information loss is prevented.\\

\begin{figure}
\centering
\includegraphics[width=0.5\linewidth]{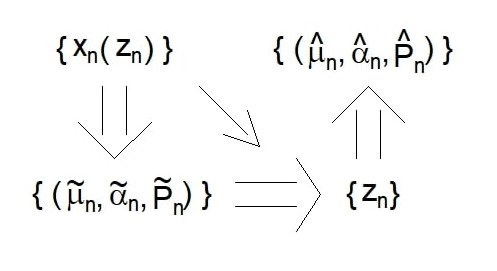}
\caption{{Illustration of the process of estimation of \geng\ model parameters. For a given data $\{x_n(z_n)\}_{n=1}^N$, the first step is to make pre-estimates $\{(\tilde{\mu}_n,\tilde{\alpha}_n,\tilde{P}_n)\}_{n=1}^N$. After that, K-means clustering of the three-dimensional sequence of pre-estimates gives us $\{z_n\}_{n=1}^N$. Finally, having the sequence of the environment states, one obtains the sequence of the estimates $\{(\widehat{\mu}_n,\widehat{\alpha}_n,\widehat{P}_n)\}_{n=1}^N$ of model parameters. In the previous approach, instead of defining pre-estimates, K-means was applied directly on $\{x_n(z_n)\}_{n=1}^N$ (shown by a diagonal arrow in diagram).}}
\label{fig:illustration}
 \end{figure}

As mentioned before, our goal is not to define new estimators of model parameters, but to improve the estimation of $\{z_n\}_{n=1}^\infty$. Given sequence of so-called 'pre-estimators' $\{(\tilde{\mu}_n,\tilde{\alpha}_n,\tilde{P}_n)\}_{n=1}^N$ is just a a helpful tool to estimate $\{z_n\}_{n=1}^\infty$, and it doesn't represent any kind of alternative estimates of model parameters. Model parameters have already been successfully estimated in \cite{LNR2018}, and we rely on those results in evaluating our \envst\ method. {For an illustration, see Figure~\ref{fig:illustration}}.

Although it is possible to implement clustering of three-dimensional data $\{(\tilde{\mu}_n,\tilde{\alpha}_n,\tilde{P}_n)\}_{n=1}^N$, method can be improved even more, by considering trimmed (truncated) means. To that purpose, for a given sequence $a_1,a_2,\dots ,a_N$ and vector $\mathbf{c}=(c_0,c_1,\dots ,c_k)'$ let us define a function
\begin{equation}\label{1}
T(a_i,\mathbf{c})=\begin{cases}
a_i, & i\leq k \mbox{ or } i> N-k,\\
\sum_{j=i-k}^{i+k}c_{\vert j-i\vert}a_i, & k<i\leq N-k.\\
\end{cases}
\end{equation}
Elements of $\mathbf{c}$ are decreasing $c_0\geq c_1\geq\dots\geq c_k$, so that $T(a_i,\mathbf{c})$ represents a trimmed mean affected the most by the current value $a_i$. The effect of the $k$ neighboring elements of $a_i$ decreases when moving away from $a_i$. If $z_{n-k}=\dots=z_{n+k}$ for some $n\in\{1,\dots,N\}$, it will make sense to use $T(\tilde{\mu}_n,\mathbf{c})$ instead of $\tilde{\mu}_n$ when estimating the environment state $z_n$, because in this case all the elements $X_{n-k},\dots ,X_{n+k}$ carry information about $z_n$. As mentioned in \cite{LNR2018}, $RrNGINAR(\mathcal{M,A,P})$ model shows bad performances in case when environment states are changing rapidly. Its application makes sense only if the probability of remaining in the same state is big enough. Hence, for $k$ small enough, one may assume that $2k+1$ neighboring elements of $\{z_n\}_{n=1}^N$ are equal with high probability in all the situations of practical importance. Thus, it sounds reasonable to replace $\{(\tilde{\mu}_n,\tilde{\alpha}_n,\tilde{P}_n)\}_{n=1}^N$ in clustering procedure with another three-dimensional data sequence $\{(T(\tilde{\mu}_n,\mathbf{c}_m), T(\tilde{\alpha}_n,\mathbf{c}_a),T(\tilde{P}_n,\mathbf{c}_p))\}_{n=1}^N$, for some vectors $\mathbf{c}_m$, $\mathbf{c}_a$ and $\mathbf{c}_p$. Theoretically speaking, lengths of these vectors do not have to be equal. The upper limit of the vector's length $k$ might be discussed as well. Higher values of $k$ give better pre-estimates, provided all of observations $X_{n-k},\dots ,X_{n+k}$ correspond to the same state. Otherwise, pre-estimates might be even worsened.\ 
To reconcile these two opposing facts, we are not going to discuss vectors $\mathbf{c}_m$, $\mathbf{c}_a$, $\mathbf{c}_p$ of length higher than $4$.\\

If we want all the coordinates of $(T(\tilde{\mu}_n,\mathbf{c}_m), T(\tilde{\alpha}_n,\mathbf{c}_a),T(\tilde{P}_n,\mathbf{c}_p))$ to have equal impact on the clustering, it is necessary to scale them. Thus, we define a function that to a given element $a_n$ of a sequence $\{a_n\}_{n=1}^N$ assigns properly scaled (normed) value of $T(a_n,\mathbf{c})$ given with

\[S(a_n,\mathbf{c})=\frac{T(a_n,\mathbf{c})\cdot N}{\sum_{i=1}^N T(a_i,\mathbf{c})}.\]

By introducing three more parameters $C_m,C_a,C_p \in \mathbb{R}$, it becomes possible to control the level of impact each coordinate has on the clustering procedure. Finally, by using standard K-means we cluster the three-dimensional data vector
\begin{equation}\label{pet}
(C_mS(\tilde{\mu}_n,\mathbf{c}_m),C_aS(\tilde{\alpha}_n,\mathbf{c}_a),C_pS(\tilde{P}_n,\mathbf{c}_p)).
\end{equation}

The only left is to define starting pre-estimators $(\tilde{\mu}_n,\tilde{\alpha}_n,\tilde{P}_n),\ n=1,2,\ldots,N$, in a reasonable way, by looking at the construction of \geng\ models. Bearing in mind the fact that parameters ${\mu_i},\ i=1,2,\ldots,r,$ represent means within clusters, it would be reasonable to set for any $n=1,2,\ldots,N$ that \begin{equation}\label{mu}{\tilde{\mu}_n}=X_n.\end{equation}

Taking into account the fact that the partial auto-correlation function is used to determine the order of the time series, estimation of the sequence $\{{P}_n\}_{n=1}^N$ will take place as follows. If $p_{z_n}$ is the maximal order allowed for particular element in the state $z_n$, than we have
\begin{equation}\label{order}
\tilde{P}_n=\begin{cases}
\max\limits_{K=1,\ldots,p_{z_n}}pacf_K(X_1,\dots X_{2d_p+1}), & n\leq d_p,\\
\max\limits_{K=1,\ldots,p_{z_n}}pacf_K(X_{n-d_p},\dots ,X_{n+d_p}), & d_p<n\leq N-d_p,\\
\max\limits_{K=1,\ldots,p_{z_n}}pacf_K(X_{N-2d_p},\dots ,X_N), & n> N-d_p,
\end{cases}
\end{equation}
where $pacf_K$ is the partial auto-correlation function at lag $K$ and $d_p\in\mathbb{N}$. The function ~\eqref{order} works well if all $2d_p+1$ elements of the sequence $\{X_n\}_{n=1}^N$  involved in $\tilde{P}_n$ correspond to the same state $z_n$. However, this requirement is not demanding, since the application of $RrNGINAR(\mathcal{M,A,P})$ model is reasonable only in the case when the probability of remaining in the same state is higher than the probability of changing the state.\\ 

To predict the thinning parameter value in moment $n$, we use the known property of the negative binomial thinning operator, that $E\left(\alpha\ast X\vert X\right)=\alpha X$. This motivates us, by looking at~\eqref{model definition}, to define

\begin{equation}\nonumber
{\alpha}_n^{*} \stackrel{}=\begin{cases}
A_n/B_n,\ &  B_n\neq 0, \quad \ n > 1,\\
1,\ & A_n=B_n=0,\ n>1,\\
 \max\left\{\left(\frac{A_l}{B_l}\colon l\in\{2,\dots,N\}, B_l>0 \right)\right\},\ & otherwise,
\end{cases}
\end{equation}
for any $n\in\mathbb{N}$, where $A_n=(x_n-T( \tilde{\mu}_n,c_m  ))_+$ and $B_n=\frac{1}{s}\sum_{i=1}^{s}A_i$ for $s=\min\{n-1,\tilde{P}_n\}$. Here $(x)_+=\max\{x,0\}$ represents the positive part of $x\in\mathbb{R}$. Since such obtained thinning parameter value might be greater than $1$, we finally have

\begin{equation}\label{alpha}
\tilde{\alpha}_n=\frac{{\alpha}_n^{*}}{\max\limits_{n=1,\ldots,N}{\alpha}_n^{*}},\ \ n\in N.
\end{equation}

To apply new \envst\  method, one should choose the values of parameters $d_p$, $\mathbf{c}_m$, $\mathbf{c}_a$, $\mathbf{c}_p$, $C_m$, $C_a$ and $C_p$ (called in sequel \envst\ method parameters). We will try to make an optimal choice based on $RrNGINAR_{max}(\mathcal{M,A,P})$ and $RrNGINAR_{1}(\mathcal{M,A,P})$ simulations. It is important here to distinguish  \envst\ method parameters from the model parameters. In Section~\ref{sec1} we give details about our choice of model parameters, while in Section \ref{sec2} we give results of the simulation study with such choice of model parameters and discuss how to choose \envst\ method parameters.\\

\section{Simulation study---the choice of model parameters}\label{sec1}

We simulated $RrNGINAR(\mathcal{M,A,P})$ time series of length $N=500$. In the sequel we consider all sequences of the same length, so instead of $\{\cdot\}_{n=1}^{500}$ we write shortly $\{\cdot\}$. Properties of simulations are such that they make difficult to apply standard K-means.
The case with $r=2$ diffe\-rent environment states is presented within the section. On the other hand, the case with $r=3$ environment states can be found in Appendix A. For each of the cases, two different combinations of model parameters are observed. Further, each combination of parameters will generate two different replications of the corresponding $RrNGINAR(\mathcal{M,A,P})$ time series. One of them will be used to obtain the values of $d_p,\mathbf{c}_m,\mathbf{c}_a,\mathbf{c}_p,C_m,C_a,C_p$. With the help of such obtained \envst\ method parameters, the other replication will be reconstructed in order to evaluate efficiency of the new \envst\ method. Furthermore, both versions of the model, $RrNGINAR_{max}(\mathcal{M,A,P})$ and $RrNGINAR_{1}(\mathcal{M,A,P})$ will be analyzed simultaneously. For more information about these models, see \cite{LNR2018}.\\

The random environment process, being a Markov chain, has parameters $p_{vec}$ --- a vector containing initial probabilities of being in certain state and $p_{mat}$ --- transition probability matrix that in the intersection of $i$-th row and $j$-th column contains the probability $P(Z_n=i\vert Z_{n-1}=j)$ for any $i,j\in\{1,\dots r\}$. Another remark about the notation is that we write $\mathcal{M}$, $\mathcal{A}$ and $\mathcal{P}$ as vectors, even though they are introduced as sets. We do so to eliminate the ambiguity, preserving the order of the states.\\

In order to create $R2NGINAR(\mathcal{M,A,P})$ simulations, the following combinations of parameters are given.
\begin{enumerate}
\item  First of all, we are going to create time series with similar means within states, while other model parameters will differ significantly. Surrounding like this would make K-means useless. Hence, we choose $\mathcal{M}=(1,1.5)$. On the contrary to that, thinning parameters, as well as maximal orders within states, should differ significantly. Hence, we choose $\mathcal{A}=(0.05,0.6)$ and $\mathcal{P}=(2,4)$. Regarding the choice of $\alpha_j,\ j=1,2,$ one of them is chosen to be very small, while the other is chosen to be close to its upper limit. Furthermore, probabilities $\phi_{i,j}^k$ corresponding to the \gengmTwo\ simulation are chosen to be
$$\phi_1=\left[\begin{array}{cc}
1 & 0\\
0.9 & 0.1
\end{array}\right],\ \ \phi_2=\left[\begin{array}{cccc}
1 & 0 & 0 & 0\\
0.1 & 0.9 & 0 & 0\\
0.1 & 0.45 & 0.45 & 0\\
0.1 & 0.1 & 0.4 & 0.4
\end{array}\right].$$
 Probabilities corresponding to the \gengoTwo\ simulation are located in last rows of these matrices. An initial state is nearly fair, due to the value of its distribution $p_{vec}=(0.6,0.4)$. In order to have long arrays of elements corresponding to the same state within the simulated $R2NGINAR(\mathcal{M,A,P})$ time series, transition probabilities outside the main diagonal are significantly smaller than those located on the main diagonal. Thus, transition probability matrix is of the form  \[p_{mat}=\left[\begin{array}{cc}
0.9 & 0.1\\
0.2 & 0.8
\end{array}\right].\]

\item The other combination of parameters is characterized by a great similarity between thinning parameters. Beside that, the mean values will be similar enough to to make it difficult to use the standard K-means method. Orders of the model will be the only values on the basis of which it is possible to determine the environment states of realizations. That will be a good test for our new approach. To that purpose, we have that $\mathcal{M}=(3,5)$, $\mathcal{A}=(0.4,0.5)$ and $\mathcal{P}=(2,5)$. Further, {in the case of } \gengmTwo, $$\phi_1=\left[\begin{array}{cc}
1 & 0\\
0.4 & 0.6
\end{array}\right],\ \ \phi_2=\left[\begin{array}{ccccc}
1 & 0 & 0 & 0 & 0\\
0.2 & 0.8 & 0 & 0 & 0\\
0.4 & 0.4 & 0.2 & 0 & 0\\
0.3 & 0.3 & 0.3 & 0.1 & 0\\
0,4 & 0.2 & 0.2 & 0.1 & 0.1
\end{array}\right].$$
Last rows of these matrices contain probabilities corresponding to the \gengoTwo\ simulation. An initial state is fair, since $p_{vec}=(0.5,0.5)$. Finally, the transition probability matrix provides long arrays of elements corresponding to the same state, that is, \[p_{mat}=\left[\begin{array}{cc}
0.8 & 0.2\\
0.25 & 0.75
\end{array}\right].\]

\end{enumerate}

\section{Simulation study --- results and the choice of \envst\ method parameters}\label{sec2}

In this section we seek for optimal \envst\ method parameters, based on $RrNGINAR(\mathcal{M,A,P})$ simulations with $r=2$ environment states. The choice of corresponding model parameters is given in the previous section. In order to improve the readability of the manuscript, only one parameters combination will be discussed in detail, with fully exposed procedure of obtaining $d_p$, $\mathbf{c}_m$, $\mathbf{c}_a$, $\mathbf{c}_p$, $C_m, C_a$ and $C_p$. As for the second combination, the procedure will be omitted and only final results will be provided. Corresponding discussion regarding optimal \envst\ method parameters {in the case of }$RrNGINAR(\mathcal{M,A,P})$ simulations with $r=3$ environment states is provided in Appendix B.\\

\begin{figure}
\centering
\begin{minipage}[b]{0.85\textwidth}
	 	\includegraphics[width=1\linewidth]{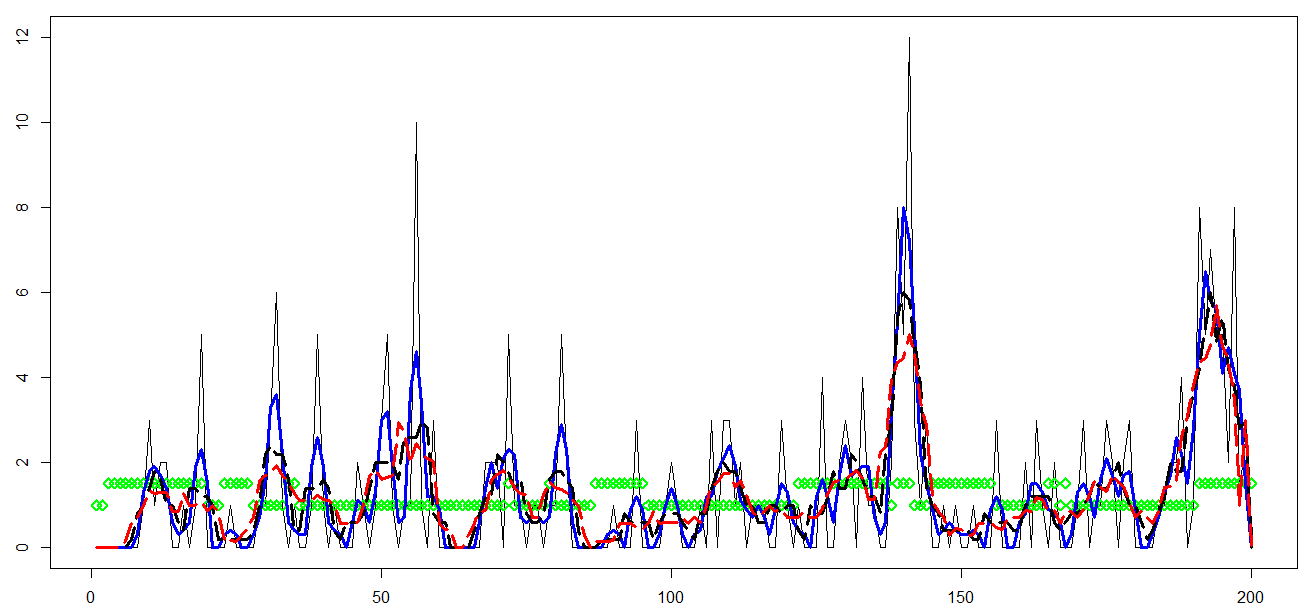}
	         	\caption*{\footnotesize a) $R2NGINAR_{max}(2,4)$ model}
 \end{minipage}
\begin{minipage}[b]{0.85\textwidth}
	 	\includegraphics[width=1\linewidth]{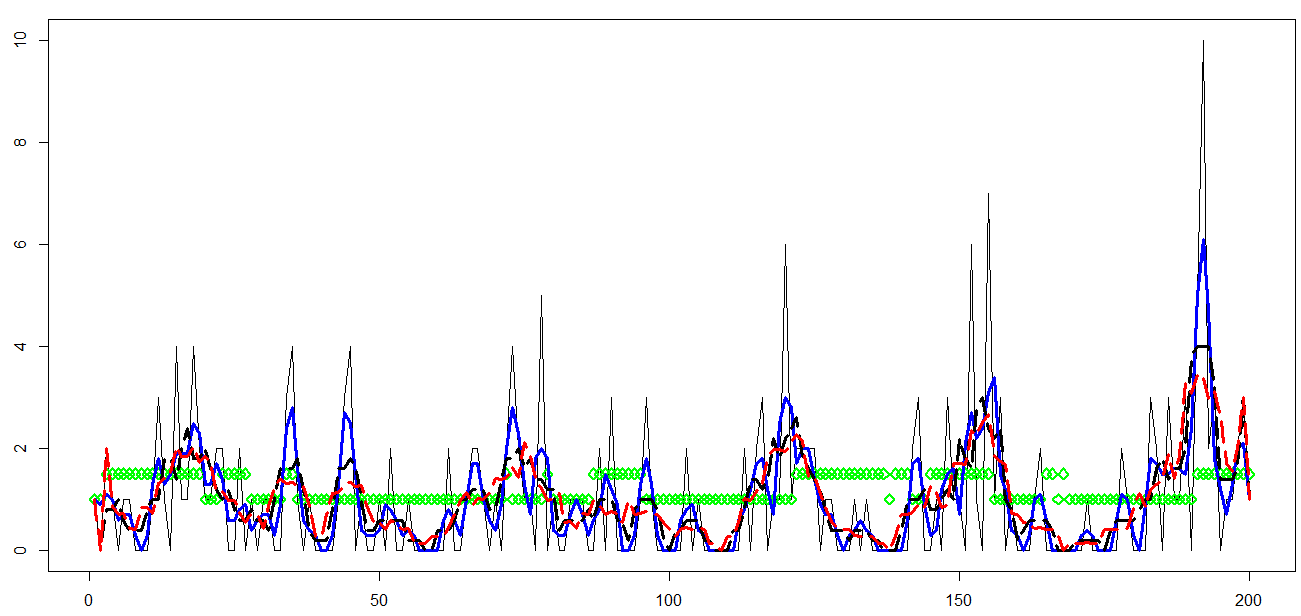}
	         	\caption*{\footnotesize b) $R2NGINAR_1(2,4)$ model}
 \end{minipage}
\caption{\footnotesize Pre-estimates of $\{{\mu_n}\}$ obtained for various selections of $\mathbf{c}_m$ in the case of simulated $R2NGINAR(2,4)$ models: green diamond-exact mean values sequence $\{\mu_n\}$; regular black line-sequence $\{T(\tilde{\mu}_n,\mathbf{c}_m)\}$ for $\mathbf{c}_m=1$; thick blue line- sequence $\{T(\tilde{\mu}_n,\mathbf{c}_m)\}$ for $\mathbf{c}_m=(0.4,0.3)$; dashed black line-sequence $\{T(\tilde{\mu}_n,\mathbf{c}_m)\}$ for $\mathbf{c}_m=(0.2,0.2,0.2)$, dashed red line-sequence $\{T(\tilde{\mu}_n,\mathbf{c}_m)\}$ for $\mathbf{c}_m=(0.16,0.14,0.14,0.14)$.}
\label{fig:first comb}
\end{figure}

Using the first combination of the model parameters, corresponding $R2NGINAR_{max}(2,4)$ and $R2NGINAR_{1}(2,4)$ simulations were created, two replications of each. The first replication of each pair was used to provide the parameters of \envst\ method. The procedure starts with the determination of $\{\tilde{\mu}_n\}$ using (\ref{mu}). In order to improve $\{{\tilde{\mu}}_n\}$, vector $\mathbf{c}_m$ has to be provided. For $k$ small enough, we have already assumed that all $X_{n-k},\ldots,X_{n+k}$ correspond to the same state. Thus, all ${\tilde{\mu}}_{n-k},\ldots,\tilde{\mu}_{n+k}$ can have similar contribution to $T(\tilde{\mu}_n,\mathbf{c}_m)$. In other words, we can choose  coordinates of the vector $\mathbf{c}_m$ to be as equal as possible. Due to the fact that it multiplies the middle realization $x_n$, the value of $c_0$ may eventually be a bit higher. Figure~\ref{fig:first comb} shows sequences of pre-estimates $\{T(\tilde{\mu}_n,\mathbf{c}_m)\}$, obtained for various selections of $\mathbf{c}_m$. There we show only the first $200$ elements, to increase readability of the plot.\\

As Figure~\ref{fig:first comb} shows, the usage of $\mathbf{c}_m$ results in much more accurate pre-estimates of the sequence $\{\mu_n\}$ in both cases. Using this technique, we managed to trim peaks that deviate significantly form the real mean values. Obviously, the best result is obtained for $\mathbf{c}_m=(0.16,0.14,0.14,0.14)$ in the case of $R2NGINAR_{max}(2,4)$ simulation. Speaking of $R1NGINAR_{1}(2,4)$ simulation, similar results are obtained in the case of $\mathbf{c}_m=(0.2,0.2,0.2)$ and $\mathbf{c}_m=(0.16,0.14,0.14,0.14)$. The second option is selected.\\

We determine $\{\tilde{P_n}\}$ in two steps. The first step provides a dete\-rmination of the parameter $d_p$, given in (\ref{order}). In order to obtain the optimal value of $d_p$, let us denote by $\Delta_p$ the root mean square of differences between correct orders $P_n,\ n=1,2,\ldots,500,$ and corresponding estimated order values $\tilde{P}_n,\ 1,2,\ldots,500,$ obtained by (\ref{order}). The error $\Delta_p$ is calculated for various choices of $d_p$ and the results are presented in Table~\ref{tabla1}. The smallest value of $\Delta_p$ will reveal the optimal value of parameter $d_p$. Having a brief look at Table~\ref{tabla1}, we conclude that, in the case of $R2NGINAR_{max}(2,4)$ simulation, the smallest value of $\Delta_p$ is obtained for $d_p=8\ (\Delta_p=1.439)$. Similarly, the smallest $\Delta_p$ value in the case of $R2NGINAR_{1}(2,4)$ simulation is obtained for $d_p=15\ (\Delta_p=1.372)$.\\

\begin{table}[htbp]
{\caption{\footnotesize {Values of the error $\Delta_p$} for various selections of $d_p$}\label{tabla1}{
{\scriptsize\vskip 3mm\begin{center}  \begin{tabular}{|c|c|c|c|c|c|c|c|}\hline

 \multicolumn{2}{c}{ $R2NGINAR_{max}(2,4)$   }\vline & \multicolumn{2}{c}{$R2NGINAR_{1}(2,4)$} \vline& \multicolumn{2}{c}{ $R2NGINAR_{max}(2,5)$   }\vline & \multicolumn{2}{c}{ $R2NGINAR_{1}(2,5)$    } \\
\hline
 $d_p$ & $\Delta_p$ &  $d_p$ & $\Delta_p$ & $d_p$ & $\Delta_p$ &  $d_p$ & $\Delta_p$ \\
\hline
 5 & 1.479 &  5 & 1.561 & 5 & 2.034 & 5 & 2.127 \\
 6 & 1.457 &  6 & 1.582 & 6 & 2.139 & 6 & 2.050 \\
 7 & 1.451 &  7 & 1.589 & 7 & 2.220 & 7 & 2.054 \\
 \textbf{8} & \textbf{1.439} &  8 & 1.613 & 8 & 2.166 & 8 & 2.072 \\
 9 & 1.481 &  9 & 1.621 & 9 & 2.172 & \textbf{9} & \textbf{2.010} \\
 10 & 1.519 &  10 & 1.522 & 10 & 2.110 & 10 & 2.069 \\
 11 & 1.504 &  11 & 1.493 & 11 & 2.124 & 11 & 2.099 \\
 12 & 1.457 &  12 & 1.511 & 12 & 2.085 & 12 & 2.101 \\
 13 & 1.476 &  13 & 1.496 & 13 & 2.083 & 13 & 2.089 \\
 14 & 1.483 &  14 & 1.431 & 14 & 2.035 & 14 & 2.075 \\
 15 & 1.475 &  \textbf{15} & \textbf{1.372} & 15 & 2.078 & 15 & 2.104 \\
 16 & 1.496 &  16 & 1.425 & 16 & 2.042 & 16 & 2.100 \\
 17 & 1.513 &  17 & 1.391 & \textbf{17} & \textbf{2.016} & 17 & 2.138 \\
 18 & 1.442 &  18 & 1.373 & 18 & 2.052 & 18 & 2.121 \\
 19 & 1.458 &  19 & 1.381 & 19 & 2.058 & 19 & 2.089 \\
 20 & 1.441 &  20 & 1.447 & 20 & 2.065 & 20 & 2.090 \\
\hline
\end{tabular}
\end{center}}}}
\end{table}

\begin{figure}
\centering
\begin{minipage}[b]{0.85\textwidth}
	 	\includegraphics[width=1\linewidth]{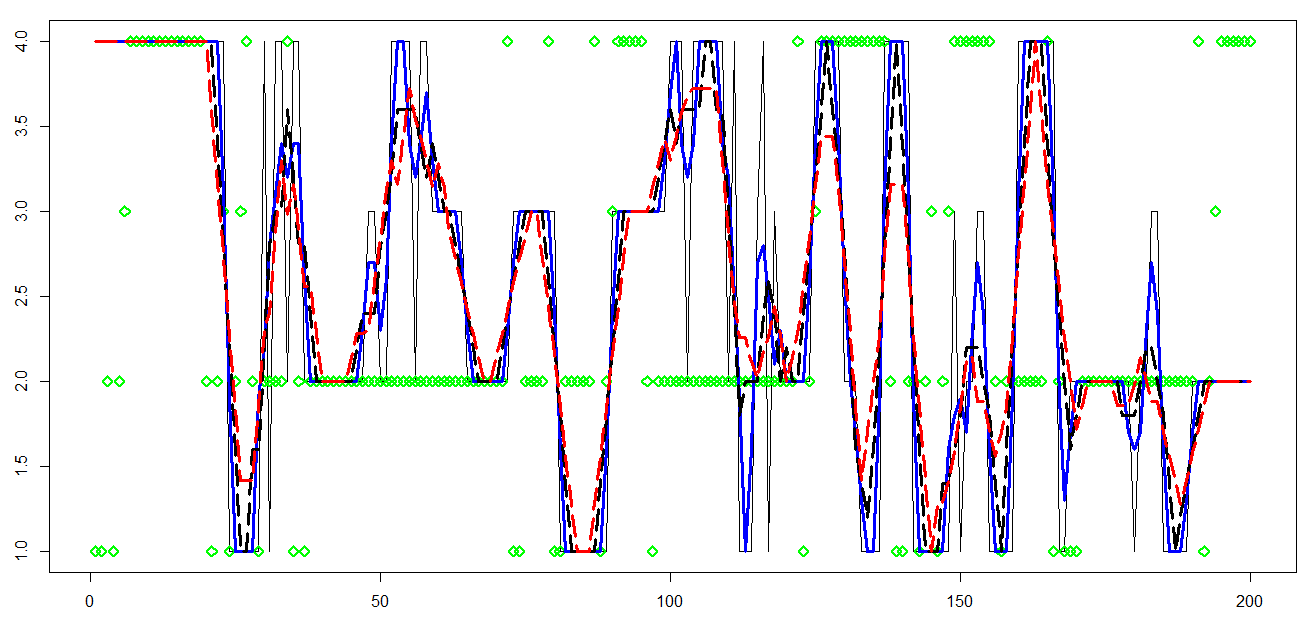}
	         	\caption*{\footnotesize a) $R2NGINAR_{max}(2,4)$ model with $d_p=8$}
 		
 \end{minipage}
\begin{minipage}[b]{0.85\textwidth}
	 	\includegraphics[width=1\linewidth]{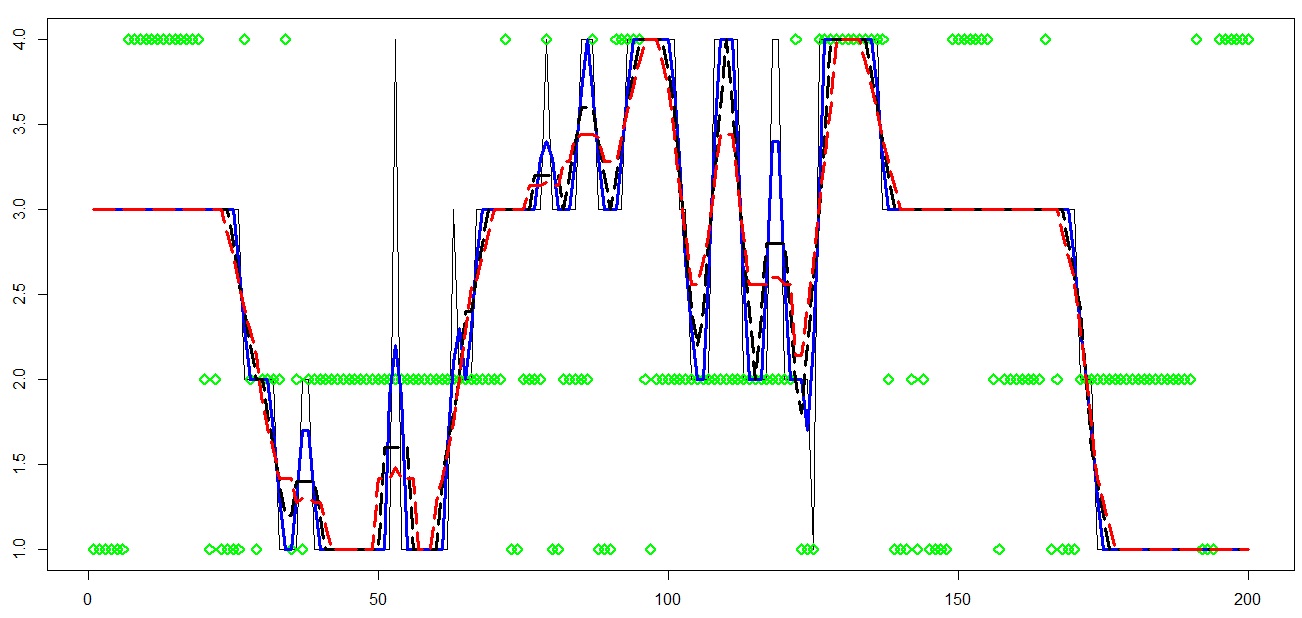}
	         	\caption*{\footnotesize b) $R2NGINAR_{1}(2,4)$ model with $d_p=15$}
 		\end{minipage}
		\caption{\footnotesize Pre-estimates of $\{P_n\}$ obtained for various selections of $\mathbf{c}_p$ in the case of simulated $R2NGINAR(2,4)$ models: green diamond-exact order sequence $\{P_n\}$; regular black line-sequence $\{T(\tilde{P}_n,\mathbf{c}_p)\}$ for $\mathbf{c}_p=1$; thick blue line-sequence $\{T(\tilde{P}_n,\mathbf{c}_p)\}$ for $\mathbf{c}_p=(0.4,0.3)$; dashed black line-sequence $\{T(\tilde{P}_n,\mathbf{c}_p)\}$ for $\mathbf{c}_p=(0.2,0.2,0.2)$, dashed red line-sequence $\{T(\tilde{P}_n,\mathbf{c}_p)\}$ for $\mathbf{c}_p=(0.16,0.14,0.14,0.14)$.}
		\label{fig:cp}
 \end{figure}

The second step involves determination of the corresponding ve\-ctor $\mathbf{c}_p$ for fixed optimal value of $d_p$. Similarly as for $c_m$, we assume that $X_{n-k},\ldots,X_{n+k}$ all correspond to the same state. Thus, ${\tilde{P}}_{n-k},\ldots,\tilde{P}_{n+k}$ are all assumed to have similar contribution to $T(\tilde{P}_n,\mathbf{c}_p)$. Hence, coordinates of $\mathbf{c}_p$ are chosen to be as similar as possible. Sequences $\{T(\tilde{P}_n,\mathbf{c}_p)\}$ obtained for various selections of $\mathbf{c}_p$ are shown in Figure~\ref{fig:cp} and compared to the exact order sequence $\{P_n\}$.\\

In order to interpret Figure~\ref{fig:cp}, one fact needs to be clarified. Namely, the goal is to choose order pre-estimate which provides the highest probability of placing corresponding observations in correct clusters. In other words, the best pre-estimate of $\{P_n\}$ is not necessarily the one that most often matches the exact order value, but the one that is close enough in most of the cases. In this respect, the best result in both cases ($R2NGINAR_{max}(2,4)$ and $R2NGINAR_{1}(2,4)$) is obtained for $k=4$, ie. for $\mathbf{c}_p=(0.16,0.14,0.14,0.14)$. Although this pre-estimate struggle to reach maximal orders, in most of the cases it stays close enough to the correct order values and do not make large mistakes.\\

\begin{figure}
\centering
\begin{minipage}[b]{0.85\textwidth}
	 	\includegraphics[width=1\linewidth]{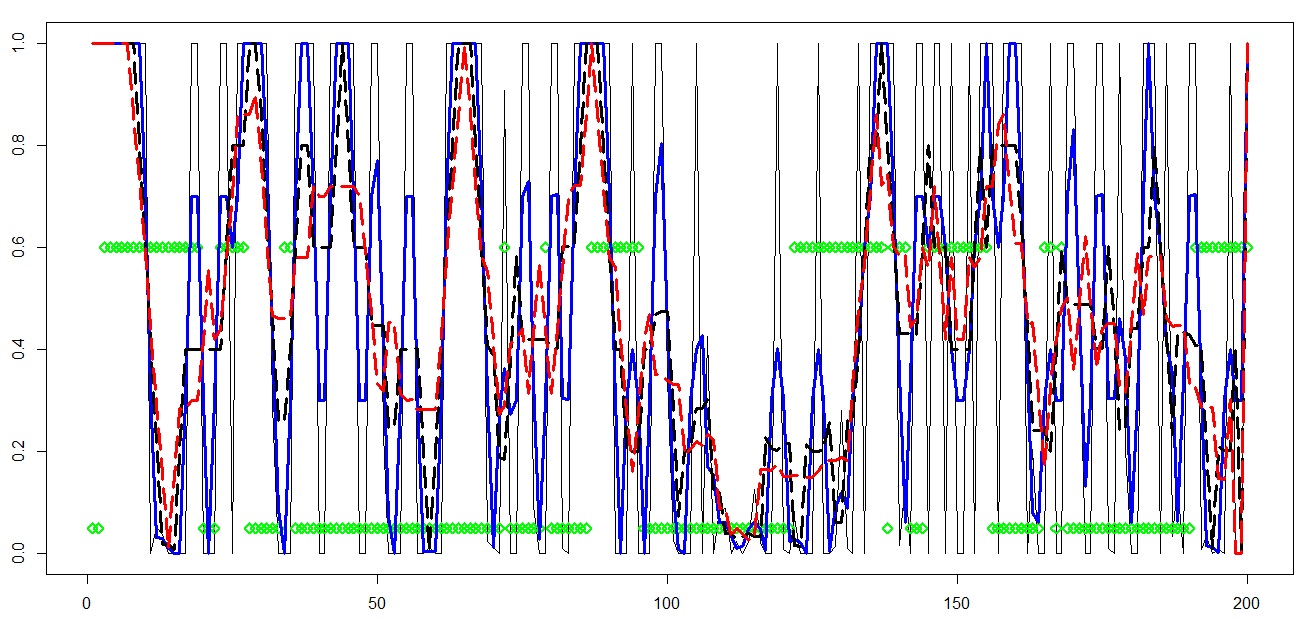}
	         	\caption*{\footnotesize a) $R2NGINAR_{max}(2,4)$ model}
 		\label{slika5}
 \end{minipage}
\begin{minipage}[b]{0.85\textwidth}
	 	\includegraphics[width=1\linewidth]{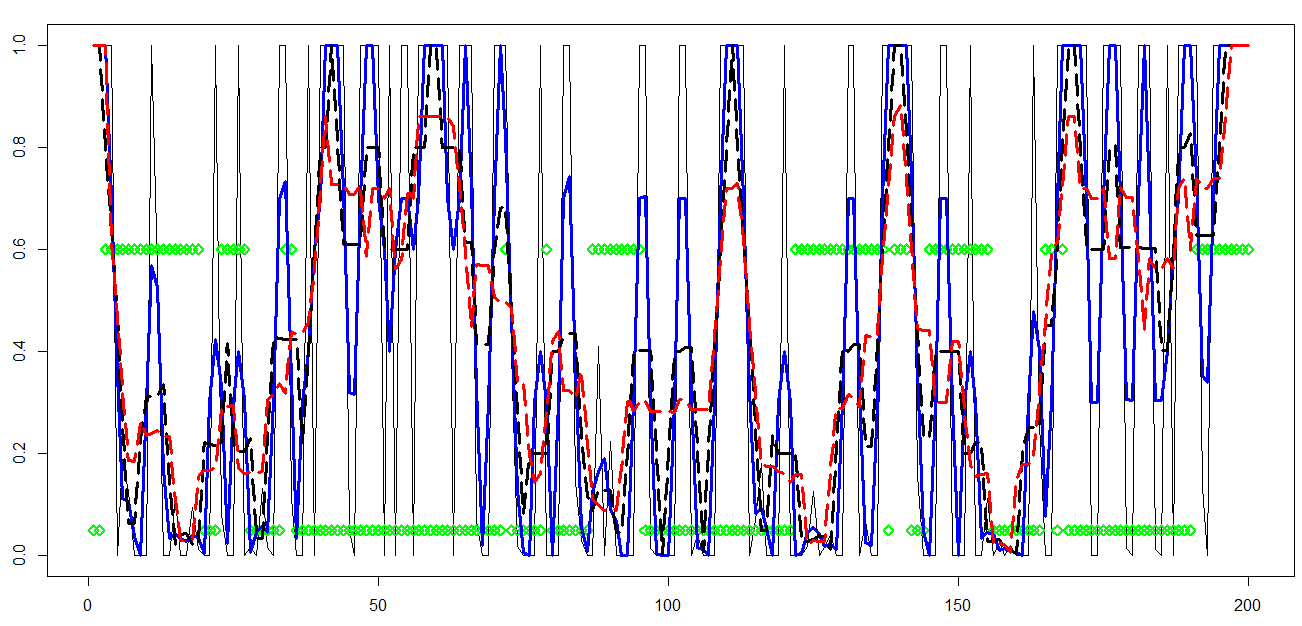}
	         	\caption*{\footnotesize b) $R2NGINAR_{1}(2,4)$ model}
 		\label{slika6}
	\end{minipage}
	\caption{\footnotesize Pre-estimates of $\{\alpha_n\}$ obtained for various selections of $\mathbf{c}_a$ in the case of simulated $R2NGINAR(2,4)$ models: green diamond-exact thinning parameters sequence $\{\alpha_n\}$; regular black line-sequence $\{T(\tilde{\alpha}_n,\mathbf{c}_a)\}$ for $\mathbf{c}_a=1$; thick blue line-sequence $\{T(\tilde{\alpha}_n,\mathbf{c}_a)\}$ for $\mathbf{c}_a=(0.4,0.3)$; dashed black line-sequence $\{T(\tilde{\alpha}_n,\mathbf{c}_a)\}$ for $\mathbf{c}_a=(0.2,0.2,0.2)$, dashed red line-sequence $\{T(\tilde{\alpha}_n,\mathbf{c}_a)\}$ for $\mathbf{c}_a=(0.16,0.14,0.14,0.14)$.}
	\label{fig:cp1}
 \end{figure}

Finally, having calculated $\{T(\tilde{\mu}_n,\mathbf{c}_m)\}$ and $\{\tilde{P}_n\}$, we are able to calculate $\tilde{\alpha}_n,\ n=1,2,\ldots,N,$ using (\ref{alpha}). 
Following the same reasons as before, coordinates of $\mathbf{c}_a$ are assumed to be as similar as possible. Regarding the length of $\mathbf{c}_a$, several options are tested. Sequences $\{T(\tilde{\alpha}_n,\mathbf{c}_a)\}$ obtained for various selections of $\mathbf{c}_a$ are given in Figure~\ref{fig:cp1} and compared to the real sequence $\{\alpha_n\}$.\\

According to the figure, vectors $\mathbf{c}_a=(0.2,0.2,0.2)$ and $\mathbf{c}_a=(0.16,0.14,0.14,0.14)$ provide more accurate pre-estimates then $\mathbf{c}_a=1$ or $\mathbf{c}_a=(0.4,0.3)$. Namely, sequences obtained for $k=3$ and $k=4$ do not show sudden and sharp ups and downs so often. The most of their values stay in a strip between $\alpha_1$ and $\alpha_2$, which is an expected behavior for a fine sequence of pre-estimates. It is hard to choose the better one, but it seems that the plot line obtained for $\mathbf{c}_a=(0.16,0.14,0.14,0.14)$ stays a bit closer to the real parameter values. This conclusion holds for both, $R2NGINAR_{max}(2,4)$ and $R2NGINAR_{1}(2,4)$ models, so the same $\mathbf{c}_a=(0.16,0.14,0.14,0.14)$ is chosen in both cases.\\

To summarize, optimal values of $d_p,$ $\mathbf{c}_m$, $\mathbf{c}_a$ and $\mathbf{c}_p$, involved in \envst\ method are provided in Table \ref{pomocna tablea1}. Note that in both cases all three vectors are chosen to be of the same length $k$, which is not surprising. Recall that $k$ depends on the probabilities of staying in the same state. If we chose smaller diagonal values of $p_{mat}$, the optimal length $k$ would certainly be smaller.\\

\begin{table}[htbp]
\caption{\footnotesize Values of the constant $d_p$ and vectors $\mathbf{c}_m$, $\mathbf{c}_a$, $\mathbf{c}_p$, {in the case of }simulated $R2NGINAR(2,4)$ time series }
\begin{center}\label{pomocna tablea1}{\footnotesize\begin{tabular}{|c|c|c|c|c|c|}
\hline
\multicolumn{4}{c}{$R2NGINAR_{max}(2,4)$}\\
\hline
$d_p$& $\mathbf{c}_m$&$\mathbf{c}_a$& $\mathbf{c}_p$\\
\hline
8&(0.16,0.14,0.14,0.14) & (0.16,0.14,0.14,0.14)& (0.16,0.14,0.14,0.14)\\
\hline
\multicolumn{4}{c}{$R2NGINAR_1(2,4)$}\\
\hline
$d_p$& $\mathbf{c}_m$&$\mathbf{c}_a$& $\mathbf{c}_p$\\
\hline
15&(0.16,0.14,0.14,0.14) & (0.16,0.14,0.14,0.14)& (0.16,0.14,0.14,0.14)\\
\hline
\end{tabular}.}
\end{center}
\end{table}

\begin{figure}
\centering
\begin{minipage}[b]{0.69\textwidth}
	 	\includegraphics[width=1\linewidth]{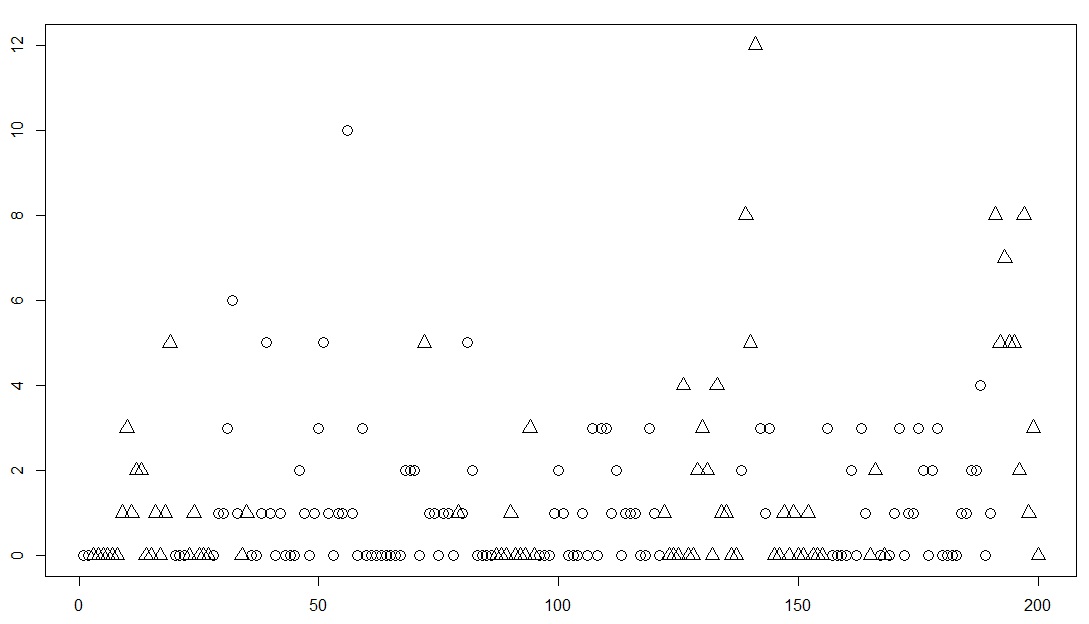}
	         	\caption*{\footnotesize Exact states of $R2NGINAR_{max}(2,4)$ simulation}
 \end{minipage}
\begin{minipage}[b]{0.69\textwidth}
	 	\includegraphics[width=1\linewidth]{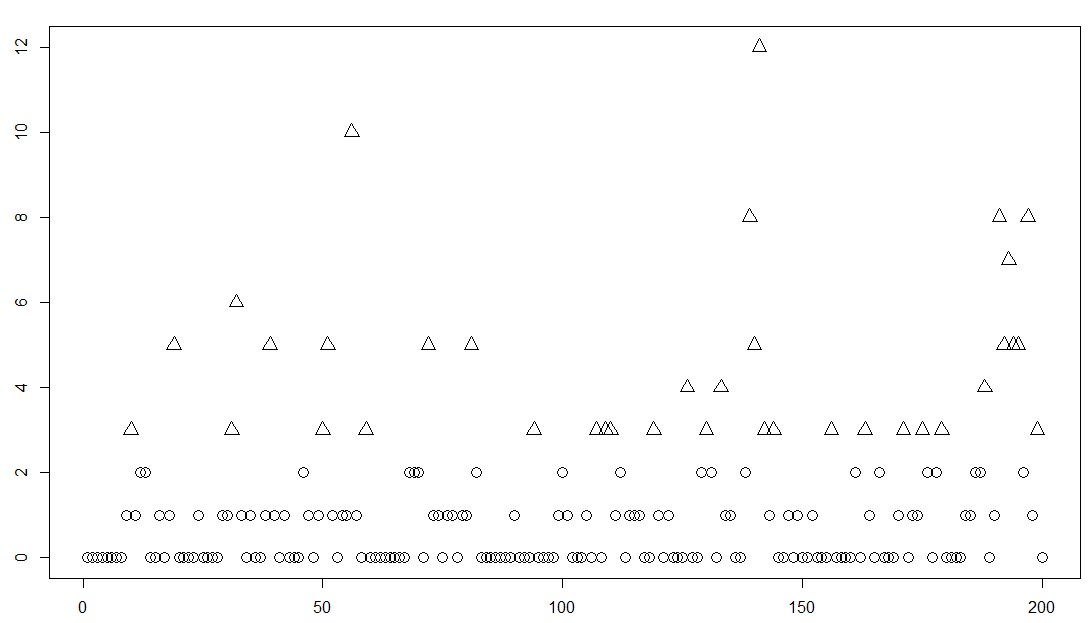}
	         	\caption*{\footnotesize States obtained by standard K-means clustering method}
 \end{minipage}
\begin{minipage}[b]{0.69\textwidth}
	 	\includegraphics[width=1\linewidth]{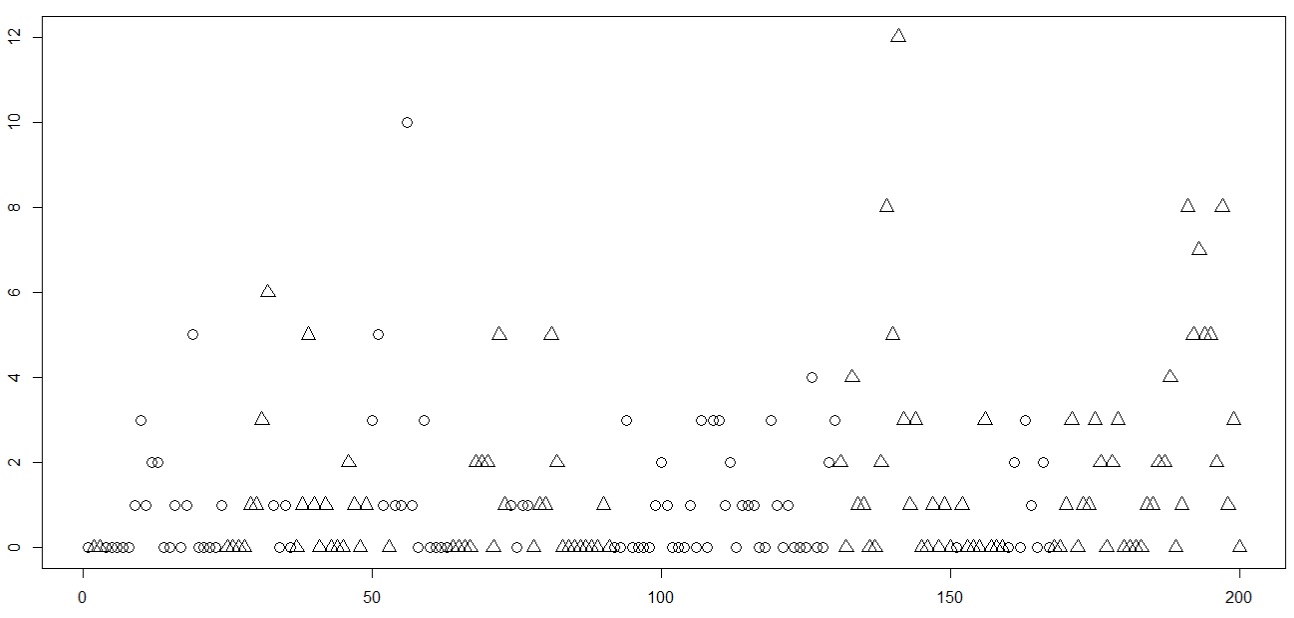}
	         	\caption*{\footnotesize States obtained by \envst\ method for $d_p=8$, $\mathbf{c}_m=(0.16,0.14,0.14,0.14)$, $\mathbf{c}_a=(0.16,0.14,0.14,0.14)$, $\mathbf{c}_p=(0.16,0.14,0.14,0.14)$, $C_m=6$, $C_a=2$, $C_p=9$.}
 \end{minipage}
\caption{\footnotesize The environment states of $R2NGINAR_{max}(2,4)$ simulation}
\label{fig:renevMax}
\end{figure}

Now, one can provide $3$-dimensional sequences  $\{T(\tilde{\mu}_n,\mathbf{c}_m)\}, \{T(\tilde{\alpha}_n,\mathbf{c}_a)\}, \{T(\tilde{P}_n,\mathbf{c}_p)\}$, and after that $\{S(\tilde{\mu}_n,\mathbf{c}_m)\}$, $\{S(\tilde{\alpha}_n,\mathbf{c}_a)\}$, $\{S(\tilde{P}_n,\mathbf{c}_p)\}$. It is left to determine parameters $C_m,C_a$ and $C_p$ given in (\ref{pet}). To that purpose, a modification of the procedure used to determine $d_p$ is applied. More precise, for each $C_m=i$, $C_a=j$, $C_p=l$, $i,j,l=1,2,\ldots, 10,$ the clustering of the three dimensional data sequence
\begin{equation}\nonumber
\{(C_mS(\tilde{\mu}_n,\mathbf{c}_m),C_aS(\tilde{\alpha}_n,\mathbf{c}_a),C_pS(\tilde{P}_n,\mathbf{c}_p))\}
\end{equation}
is performed. In that way, a thousand different estimates of the environment state sequence $\{z_n\}$ are provided. To select the best one, estimates thus obtained are compared with the sequence of exact states. The highest number of exactly estimated states will reveal the best co\-mbination of parameters $C_m$, $C_a$, $C_p$. In the case of $R2NGINAR_{max}(2,4)$ simulation, the best result in random environment estimation is obtained for $C_m=6$, $C_a=2$, $C_p=9$, having $328$ estimated states equal to corresponding exact states. On the other hand, result obtained by standard K-means managed to have $301$ exactly estimated states. A comparative overview of exact states, states obtained by standard K-means and states obtained by usage of  \envst\ method is provided by Figure~\ref{fig:renevMax}. And yet again, only first $200$ states are given in each figure.\\

Beside the higher number of the exactly estimated states, two more improvements are achieved by usage of \envst\ method. According to the figures, the \envst\ method produces much longer data series that correspond to the same state. Bearing in mind the fact that random environment models show poor performances when environment states are changing rapidly, mentioned improvement seems very convenient. Further, \envst\ method doesn't make a crisp data division using a horizontal line, as K-means does. The possibility of obtaining a high data value in the environment conditions different from those assumed for the high data values is not ruled out this time. In other words, \envst\ method allows data elements with high values to belong to the cluster with predominantly low values, and vice versa. This property makes \envst\ method more suitable for clustering the data where, beside one detected predominant environment condition, some hidden circumstances also have an impact on the time series realizations.\\

\begin{figure}
\centering
\begin{minipage}[b]{0.69\textwidth}
	 	\includegraphics[width=1\linewidth]{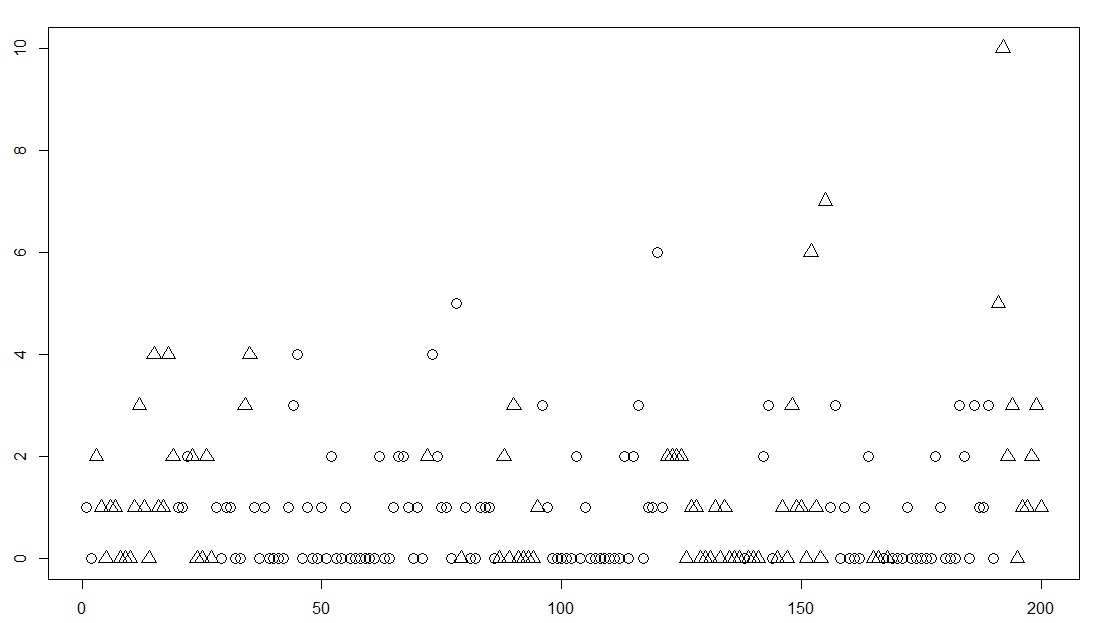}
	         	\caption*{\footnotesize Exact states of $R2NGINAR_{1}(2,4)$ simulation}
\end{minipage}
\begin{minipage}[b]{0.69\textwidth}
	 	\includegraphics[width=1\linewidth]{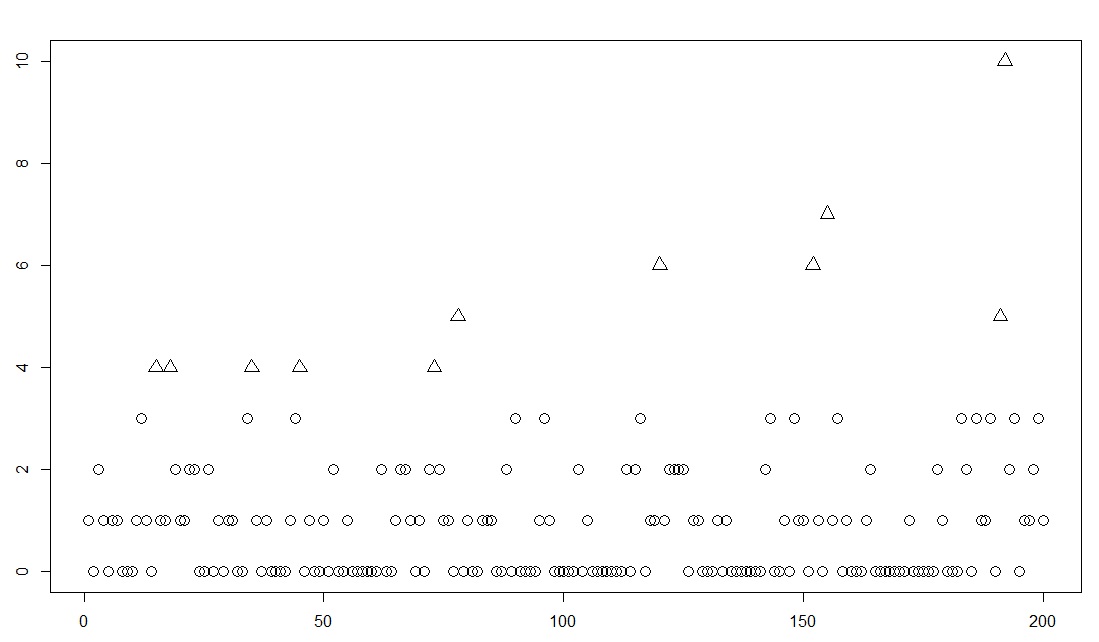}
	         	\caption*{\footnotesize States obtained by standard K-means clustering method}
  \end{minipage}
\begin{minipage}[b]{0.69\textwidth}
	 	\includegraphics[width=1\linewidth]{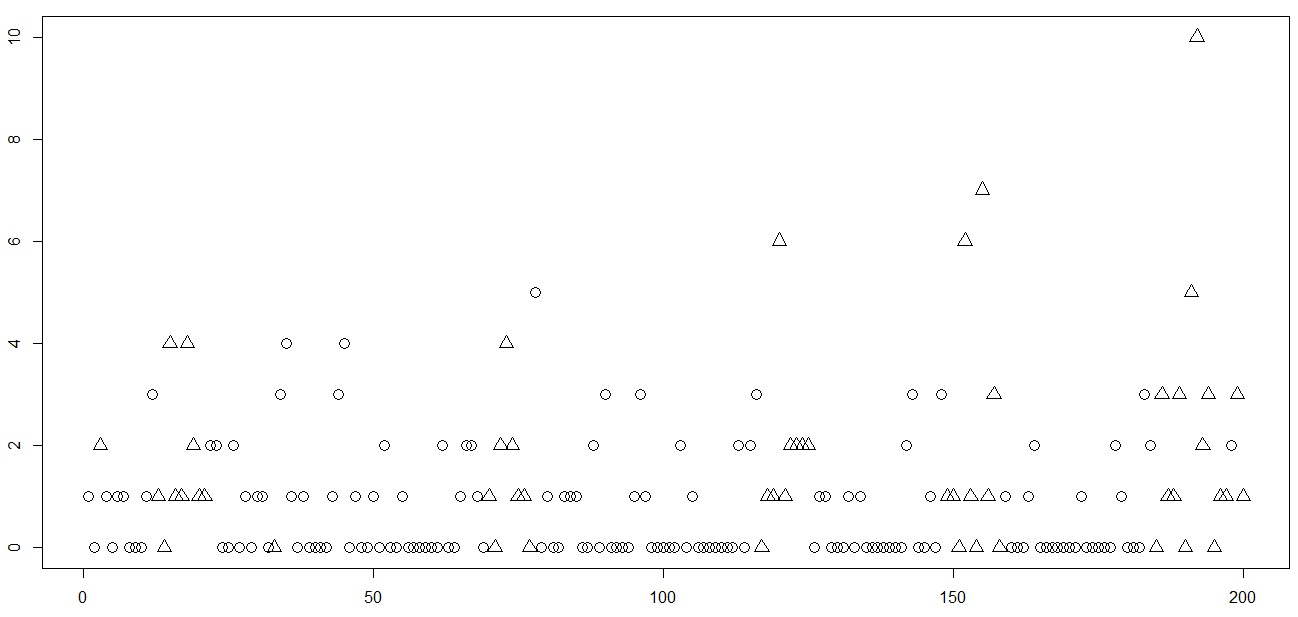}
	         	\caption*{\footnotesize States obtained by \envst\ method for $d_p=15$, $\mathbf{c}_m=(0.16,0.14,0.14,0.14)$, $\mathbf{c}_a=(0.16,0.14,0.14,0.14)$, $\mathbf{c}_p=(0.16,0.14,0.14,0.14)$, $C_m=8$, $C_a=2$, $C_p=3$.}
\end{minipage}
\caption{\footnotesize The environment states of $R2NGINAR_{1}(2,4)$ simulation}
\label{fig:renev1}
\end{figure}

Similar holds in the case of simulated $R2NGINAR_{1}(2,4)$ time series. For each $C_m=i$, $C_a=j$, $C_p=l$, $i,j,l=1,2,\ldots, 10,$ the clustering of three dimensional data sequence
\begin{equation}\nonumber
\{(C_mS(\tilde{\mu}_n,\mathbf{c}_m),C_aS(\tilde{\alpha}_n,\mathbf{c}_a),C_pS(\tilde{P}_n,\mathbf{c}_p))\}
\end{equation}
is performed. The best result is obtained for $C_m=8$, $C_a=2$, $C_p=3$, having $326$ estimated states equal to the corresponding exact states. On contrary to that, the standard K-means method managed to have $309$ exactly estimated states. And yet again, comparative overview of the exact states, states obtained by standard K-means method and states obtained by new \envst\ method is provided by Figure~\ref{fig:renev1}. The plots undoubtedly show dominance of \envst\ method comparing to the standard K-means. Achievements mentioned in the case of $R2NGINAR_{max}(2,4)$ simulation, also hold here.\\

Unused replications are suitable here to check the efficiency of our \envst\ method. First of all, these replications were observed as real-life data sequences. Further, environment state estimation via K-means method and via \envst\ method took place. The same holds for both, $R2NGINAR_{max}(2,4)$ and $R2NGINAR_{1}(2,4)$ simulations. Having results of both random environment estimation methods given above, unknown model parameters were estimated for each clustering result by usage of conditional maximum likelihood ($CML$) procedure. \\

A data sequences reconstruction by corresponding $R2NGINAR_{max}(2,4)$ or $R2NGINAR_{1}(2,4)$ model may happen now for each clustering result. $RMS$ of differences between simulated data and their reconstructions will represent the measure of the fitting quality. Results of the modeling obtained after applying standard K-means and \envst\ method are provided in Table \ref{tabela2}. Dominance of \envst\ method is noticeable. $RMS$ values obtained after applying standard K-means method are unexpecte\-dly high ($RMS=1.989$ in the case of $R2NGINAR_{max}(2,4)$ model and $RMS=1.836$ in the case of $R2NGINAR_{1}(2,4)$ model). This confirms the hypothesis given in the introduction that K-means is not {a} useful tool for clustering the data corresponding to the \gen \ process with similar means within states. On the other hand, $RMS$ values obtained after applying \envst\ method are much more acceptable ($RMS=1.529$ in the case of $R2NGINAR_{max}(2,4)$ model and $RMS=1.478$ in the case of $R2NGINAR_{1}(2,4)$ model). \\

\begin{table}[htbp]
\caption{\footnotesize $CML$ parameter estimates and $RMS$ values obtained after reconstruction of the simulated data sequences that correspond to the $R2NGINAR_{max}(2,4)$ and $R2NGINAR_{1}(2,4)$ time series}
\begin{center}\label{tabela2}{\footnotesize\begin{tabular}{|c|c|c|c|c|}
\hline
 & \multicolumn{2}{c}{ $R2NGINAR_{max}(2,4)$   }\vline &\multicolumn{2}{c}{ $R2NGINAR_1(2,4)$ }\vline\\
\hline\hline
Clustering &$CML$ & $RMS$ & $CML$ & $RMS$ \\
\hline\hline
\rule{0pt}{4ex}
Regular& $\widehat{\mathcal{M}}=(0.544,4.168)$ & 1.989 & $\widehat{\mathcal{M}}=(0.713,5.435)$ & 1.836\\
K-means & $\widehat{\mathcal{A}}=(0.001,0.403)$ & & $\widehat{\mathcal{A}}=(0.254,0.386)$ & \\
 & $\widehat{\phi}_1=\left[\begin{array}{cc}
1 & 0\\
0.999 & 0.001
\end{array}\right]$ & & $\widehat{\phi}_1=(0.937,0.063)$ & \\
 & $\widehat{\phi}_2=\left[\begin{array}{cccc}
1 & 0 & 0 & 0\\
0.001 & 0.999 & 0 & 0\\
0.328 & 0.331 & 0.341 & 0\\
0.252 & 0.200 & 0.242 & 0.306
\end{array}\right]$ & & $\widehat{\phi}_2=(0.252,0.200,0.248,0.300)$ & \\
\hline
\rule{0pt}{4ex}
 & $\widehat{\mathcal{M}}=(0.901,1.589)$ & 1.529 &$\widehat{\mathcal{M}}=(0.931,1.412)$ & 1.478 \\
 \envst & $\widehat{\mathcal{A}}=(0.002,0.309)$ & & $\widehat{\mathcal{A}}=(0.173,0.585)$ & \\
& $\widehat{\phi}_1=\left[\begin{array}{cc}
1 & 0\\
0.999 & 0.001
\end{array}\right]$ & & $\widehat{\phi}_1=(0.952,0.048)$ & \\
 & $\widehat{\phi}_2=\left[\begin{array}{cccc}
1 & 0 & 0 & 0\\
0.001 & 0.999 & 0 & 0\\
0.329 & 0.330 & 0.341 & 0\\
0.247 & 0.200 & 0.242 & 0.311
\end{array}\right]$ & & $\widehat{\phi}_2=(0.244,0.206,0.230,0.320)$ & \\
\hline\hline
\end{tabular}.}
\end{center}
\end{table}

Using the second combination of {the} model parameters, corresponding $R2NGINAR_{max}(2,5)$ and $R2NGINAR_{1}(2,5)$ simulations are created, two replications of each. {The} first one is used to obtain {the} optimal values for $d_p$, $\mathbf{c}_m$, $\mathbf{c}_a$, $\mathbf{c}_p$, $C_m$, $C_a$ and $C_p$. The same procedure as the one presented in the case of $R2NGINAR(2,4)$ simulations is preformed, and thus obtained optimal values are given in Table \ref{pomocna tabela2}.\\

\begin{table}[htbp]
\caption{\footnotesize Values of the constant $d_p$ and vectors $\mathbf{c}_m$, $\mathbf{c}_a$, $\mathbf{c}_p$, in the case of simulated $R2NGINAR(2,5)$ time series}
\begin{center}\label{pomocna tabela2}{\footnotesize\begin{tabular}{|c|c|c|c|c|c|c|}
\hline
\multicolumn{7}{c}{$R2NGINAR_{max}(2,5)$}\\
\hline
$d_p$& $\mathbf{c}_m$&$\mathbf{c}_a$& $\mathbf{c}_p$ & $C_m$ & $C_a$ & $C_p$\\
\hline
17&(0.16,0.14,0.14,0.14) & (0.16,0.14,0.14,0.14)& (0.4,0.3) & 4 & 2 & 3 \\
\hline
\multicolumn{7}{c}{$R2NGINAR_1(2,5)$}\\
\hline
$d_p$& $\mathbf{c}_m$&$\mathbf{c}_a$& $\mathbf{c}_p$ & $C_m$ & $C_a$ & $C_p$\\
\hline
9&(0.2,0.2,0.2) & (0.16,0.14,0.14,0.14)& (0.4,0.3) & 9 & 6 & 7\\
\hline
\end{tabular}}.
\end{center}
\end{table}

Further, unused replications were observed as real-life data sequences. After the environment state estimation happened via K-means method and via \envst\ method, those replications were reconstructed by $R2NGINAR_{max}(2,5)$ or $R2NGINAR_{1}(2,5)$ model for each clustering result. Modeling results thus obtained are given in Table \ref{tabela21}.\\

\begin{table}[htbp]
\caption{\footnotesize $CML$ parameter estimates and $RMS$ values obtained after reconstruction of the simulated data sequences that correspond to the $R2NGINAR_{max}(2,5)$ and $R2NGINAR_{1}(2,5)$ time series}
\begin{center}\label{tabela21}{\footnotesize\begin{tabular}{|c|c|c|c|c|}
\hline
 & \multicolumn{2}{c}{ $R2NGINAR_{max}(2,5)$   }\vline &\multicolumn{2}{c}{ $R2NGINAR_1(2,5)$ }\vline\\
\hline\hline
Clustering &$CML$ & $RMS$ & $CML$ & $RMS$ \\
\hline\hline
\rule{0pt}{4ex}
Regular& $\widehat{\mathcal{M}}=(2.480,14.170)$ & 4.030 & $\widehat{\mathcal{M}}=(2.422,13.473)$ & 4.141\\
K-means & $\widehat{\mathcal{A}}=(0.080,0.133)$ & & $\widehat{\mathcal{A}}=(0.167,0.202)$ & \\
 & $\widehat{\phi}_1=\left[\begin{array}{cc}
1 & 0\\
0.008 & 0.992
\end{array}\right]$ & & $\widehat{\phi}_1=(0.025,0.975)$ & \\
 & $\widehat{\phi}_2=\left[\begin{array}{ccccc}
1 & 0 & 0 & 0 & 0 \\
0.002 & 0.998 & 0 & 0 & 0 \\
0.399 & 0.400 & 0.201 & 0 & 0 \\
0.299 & 0.300 & 0.300 & 0.101 & 0 \\
0.198 & 0.199 & 0.200 & 0.200 & 0.203
\end{array}\right]$ & & $\widehat{\phi}_2=(0.202,0.203,0.203,0.200,0.192)$ & \\
\hline
\rule{0pt}{4ex}
 & $\widehat{\mathcal{M}}=(3.550,5.489)$ & 3.422 &$\widehat{\mathcal{M}}=(2.853,5.007)$ & 3.529 \\
 \envst & $\widehat{\mathcal{A}}=(0.010,0.433)$ & & $\widehat{\mathcal{A}}=(0.176,0.292)$ & \\
& $\widehat{\phi}_1=\left[\begin{array}{cc}
1 & 0\\
0.004 & 0.996
\end{array}\right]$ & & $\widehat{\phi}_1=(0.365,0.635)$ & \\
 & $\widehat{\phi}_2=\left[\begin{array}{ccccc}
1 & 0 & 0 & 0 & 0 \\
0.001 & 0.999 & 0 & 0 & 0 \\
0.389 & 0.399 & 0.212 & 0 & 0 \\
0.298 & 0.299 & 0.299 & 0.104 & 0 \\
0.175 & 0.201 & 0.201 & 0.200 & 0.223
\end{array}\right]$ & & $\widehat{\phi}_2=(0.176,0.200,0.202,0.221, 0.201)$ & \\
\hline\hline
\end{tabular}.}
\end{center}
\end{table}

Generally, much higher $RMS$ values were detected in the case when simulations were dictated by the second combination of model parameters. This is understandable, given that the realization values are much higher in this case. As a consequence, a benefit obtained in $RMS$ values is higher as well. Comparing the corresponding parameter estimates, we see that the estimates of means are much more accurate after application of the \envst\ method. Estimates of thinning parameters are also a bit more accurate in this case. More accurate parameter estimates ultimately led to a significant differences in $RMS$ values. Based on all the above, it can be concluded that the new \envst\ method successfully sorts the realizations in corresponding clusters and thus contributes to a more efficient application of the $R2NGINAR(\mathcal{M,A,P})$ models.

\section{Real-life data application}

\begin{figure}[h!]
	       	\centering
	 	\includegraphics[width=13cm,height=6cm]{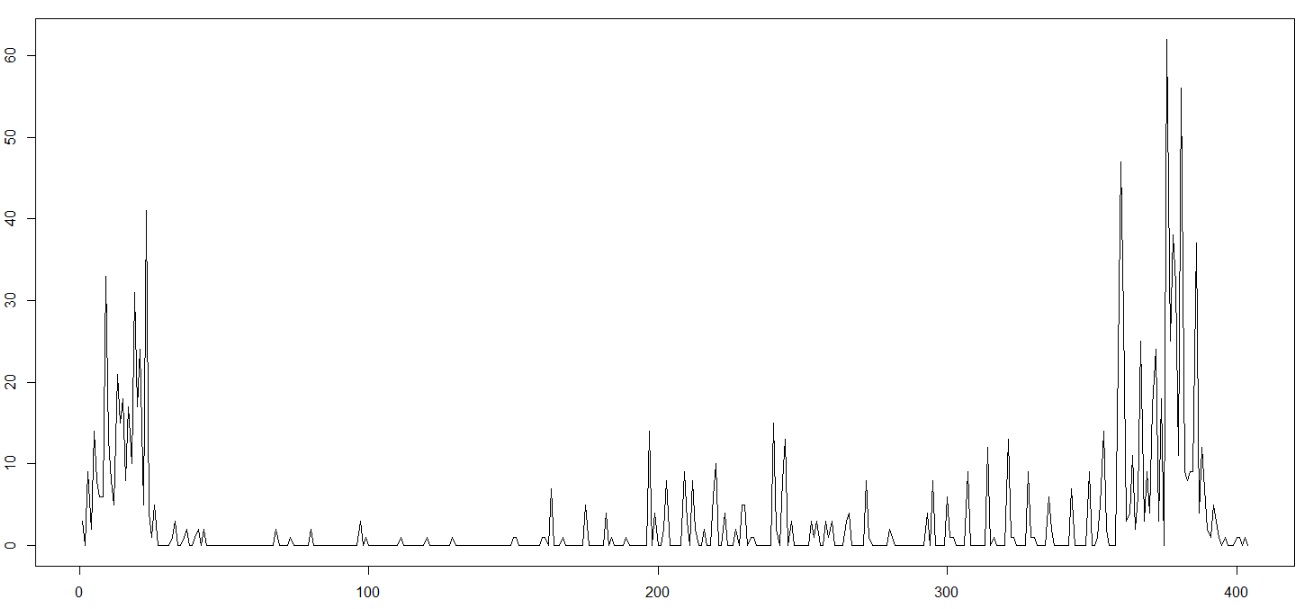}
	         	\caption{Number of newly detected COVID-19 cases in Mauritius on daily basis}
 		\label{slika29}
 \end{figure}

In order to confirm its efficiency, we tested our \envst\ method on the data that has been very popular in recent months. From the web site Data Europa (http://www.data.europa.eu) we chose the time series that represent the number of new COVID-19 cases on daily basis detected on the island of Mauritius between March 18, 2020 and April 25, 2021. The plot of a given series is provided in Figure \ref{slika29}. As can be noticed, the number of newly detected cases was kept under control most of the time. The most frequent number of newly infected inhabitants was $0$, with occasional and isolated jumps. However, in two time intervals (form March 22, 2020 to April 9, 2020 and from March 6, 2021 to April 9, 2021), strange results emerged. During those periods, the number of newly infected inhabitants oscillated dramatically, with sharp and frequent ups and downs. In other words, very high values began to appear, followed by sudden decrements and vice versa. All mentioned here indicates that environment state changes might occurred.

 \begin{figure}[h!]
	       	\centering
	 	\includegraphics[width=10cm,height=5cm]{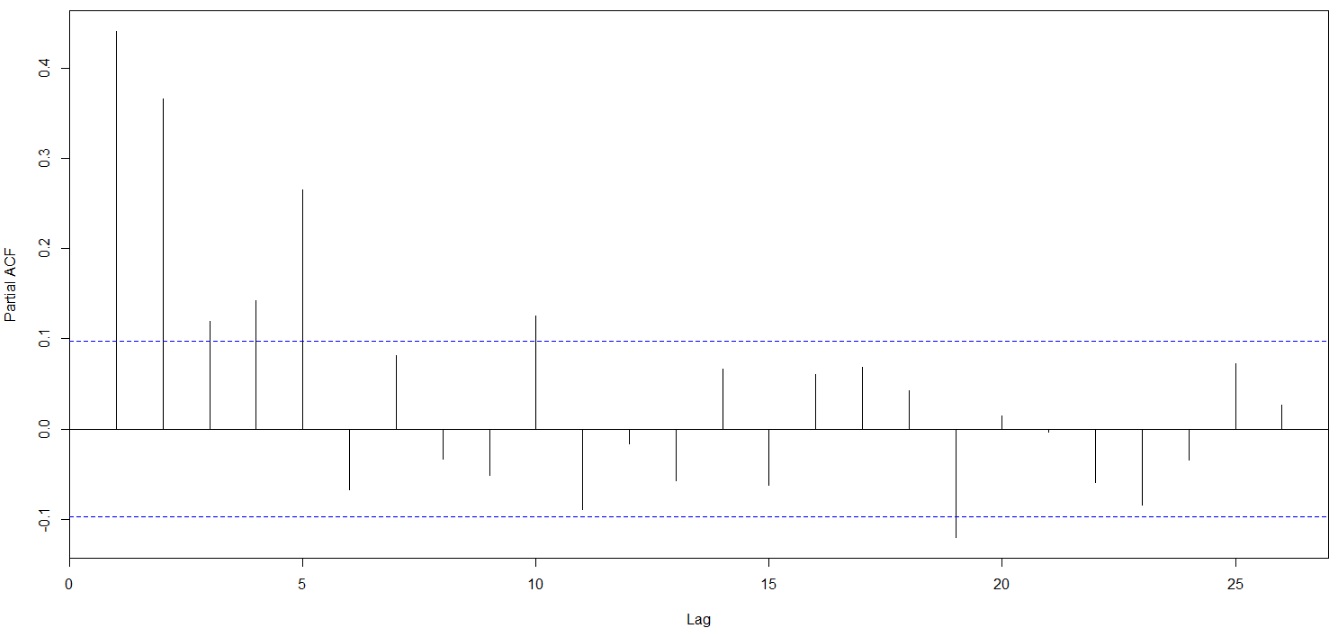}
	         	\caption{PACF for the data that represent a number of new COVID-19 cases in Mauritius on daily basis}
 		\label{slika30}
 \end{figure}

\begin{figure}
\centering
\begin{minipage}[b]{0.72\textwidth}
	 	\includegraphics[width=1\linewidth]{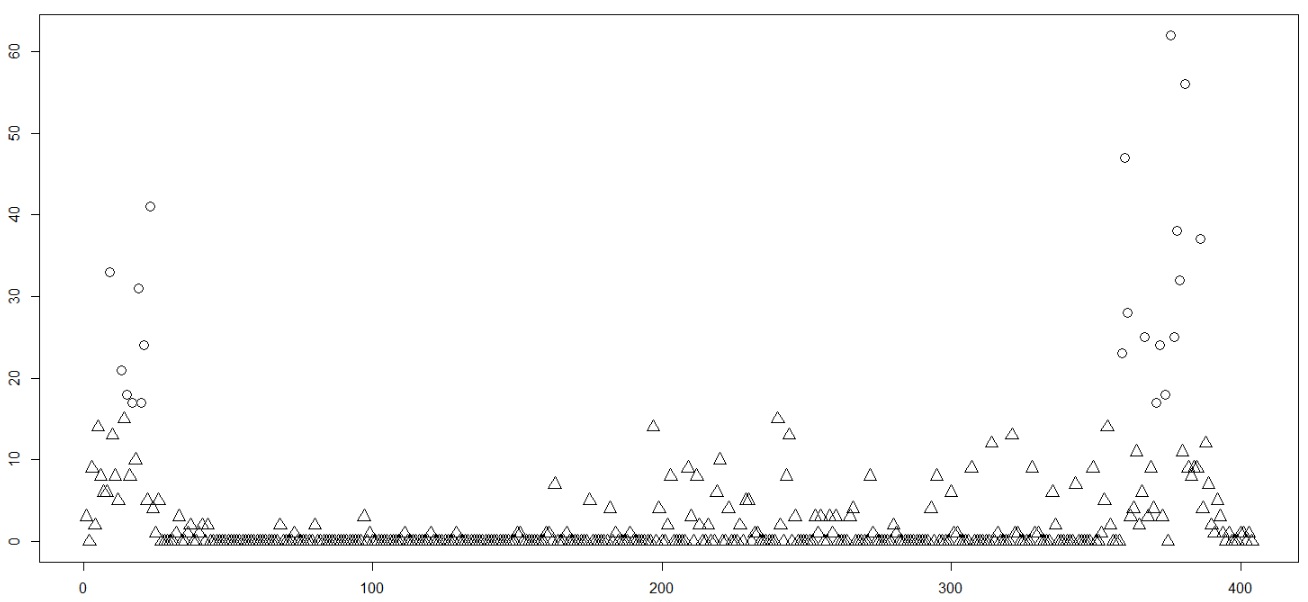}
	         	\caption*{Standard K-means}
 \end{minipage}
\begin{minipage}[b]{0.72\textwidth}
	 	\includegraphics[width=1\linewidth]{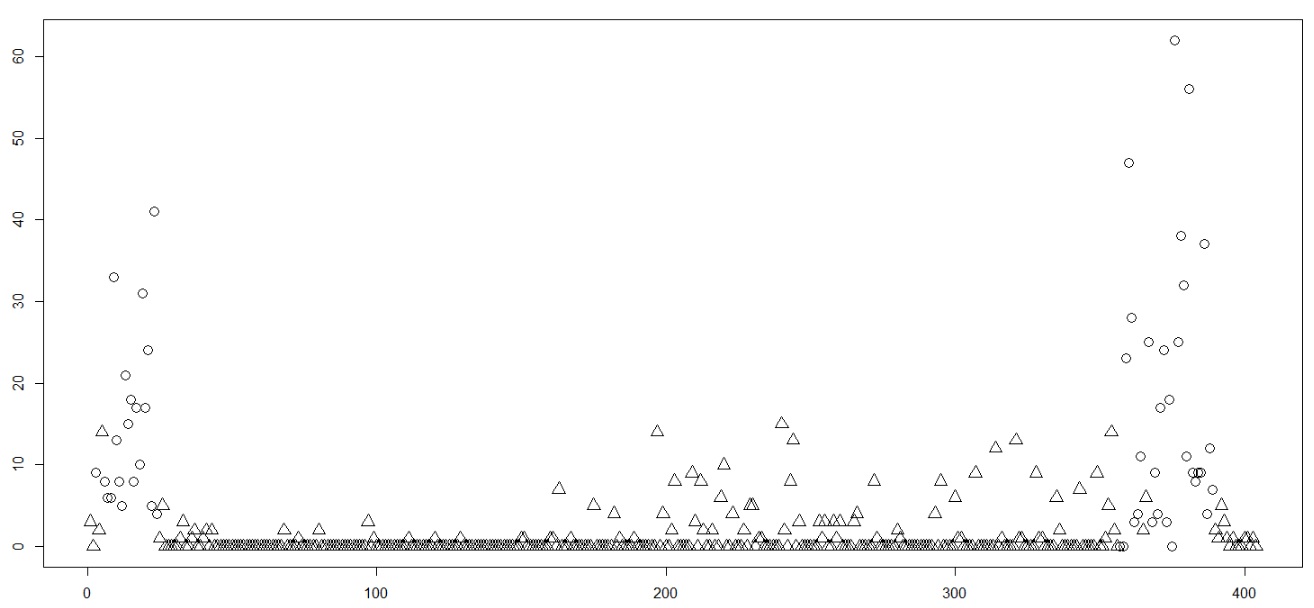}
	         	\caption*{\envst\ method}
		\end{minipage}
	\caption{Clustering results for the real-life data}	
	\label{fig:realdata}
 \end{figure}

 The plot of the autocorrelation function given in Figure \ref{slika30} shows that all orders up to order 5 are significant. Since the influence of our \envst\ method on modeling by $R2NGINAR(2,4)$ and $R2NGINAR(2,5)$ models has already been examined in {the} previous section, the same models are going to be observed here as well. Now, we can estimate random environment sequence $\{z_n\}$ using standard K-means and \envst\ method. All \envst\ method parameters are the same as in the previous section. Obtained clustering results are provided in Figure~\ref{fig:realdata}.
Unlike the standard K-means, \envst\ method had success in recognizing the atypical behavior of the time series and managed to place in a separate cluster almost all values that were realized during two mentioned time intervals. This conclusion points to the fact that the application of the selected $R2NGINAR(\mathcal{M,A,P})$ models could be even more effective after usage of \envst\ method. \\

As a final step in proving the supremacy of the \envst\ method, the fitting quality of given $R2NGINAR(2,4)$ and $R2NGINAR(2,5)$ models is examined for each clustering result. As a measure of the goodness of fit we use the root mean squares ($RMS$) of differences between the observations and their predicted values. Table \ref{tabela222} contains results obtained using $R2NGINAR_{max}(2,4)$, $R2NGINAR_{1}(2,4)$, $R2NGINAR_{max}(2,5)$ and $R2NGINAR_1(2,5)$ models for each clustering result. Obviously, there is a big difference in fitting quality, depending on the choice of method by which the realizations are distributed into clusters. According to Table \ref{tabela222}, all selected models show much lower $RMS$ values after applying the \envst\ method. In that way, usefulness of the \envst\ method is definitely proved and the benefits of its use are confirmed. If the reader possibly wants to compare results of modeling given in Table \ref{tabela222} with the results obtained using various models with stationary or non-stationary nature, he can take a look at Appendix C.\\

\begin{table}[htbp]
\caption{$CML$ parameter estimates and $RMS$-s obtained after application of two different $R2NGINAR_{max}(\mathcal{M,A,P})$ and $R2NGINAR_{1}(\mathcal{M,A,P})$ models on selected real-life data (both clustering methods are considered) }
\begin{center}\label{tabela222}{\footnotesize\begin{tabular}{|c|c|c|c|c|}
\hline\hline
 & \multicolumn{2}{c}{ $R2NGINAR_{max}(2,4)$   }\vline &\multicolumn{2}{c}{ $R2NGINAR_1(2,4)$ }\vline\\
\hline\hline
Clustering &$CML$ & $RMS$ & $CML$ & $RMS$ \\
\hline
\rule{0pt}{4ex}
Regular& $\widehat{\mathcal{M}}=(0.494,30.190)$ & 4.260 & $\widehat{\mathcal{M}}=(0.490,30.190)$ & 4.217\\
K-means & $\widehat{\mathcal{A}}=(0.001,0.481)$ & & $\widehat{\mathcal{A}}=(0.260,0.474)$ & \\
 & $\widehat{\phi}_1=\left[\begin{array}{cc}
1 & 0\\
0.001 & 0.999
\end{array}\right]$ & & $\widehat{\phi}_1=(0.019,0.981)$ & \\
 & $\widehat{\phi}_2=\left[\begin{array}{cccc}
1 & 0 & 0 & 0\\
0.001 & 0.999 & 0 & 0\\
0.329 & 0.329 & 0.342 & 0\\
0.249 & 0.199 & 0.240 & 0.312
\end{array}\right]$ & & $\widehat{\phi}_2=(0.260,0.261,0.200,0.279)$ & \\
\hline
\rule{0pt}{4ex}
 & $\widehat{\mathcal{M}}=(1.103,14.791)$ & 3.871 &$\widehat{\mathcal{M}}=(1.522,14.793)$ & 3.828 \\
 \envst & $\widehat{\mathcal{A}}=(0.001,0.511)$ & & $\widehat{\mathcal{A}}=(0.249,0.937)$ & \\
& $\widehat{\phi}_1=\left[\begin{array}{cc}
1 & 0\\
0.999 & 0.001
\end{array}\right]$ & & $\widehat{\phi}_1=(0.049,0.951)$ & \\
 & $\widehat{\phi}_2=\left[\begin{array}{cccc}
1 & 0 & 0 & 0\\
0.001 & 0.999 & 0 & 0\\
0.329 & 0.328 & 0.343 & 0\\
0.191 & 0.207 & 0.248 & 0.354
\end{array}\right]$ & & $\widehat{\phi}_2=(0.249,0.243,0.238,0.270)$ & \\
\hline\hline\hline
 & \multicolumn{2}{c}{ $R2NGINAR_{max}(2,5)$   }\vline &\multicolumn{2}{c}{ $R2NGINAR_1(2,5)$ }\vline\\
\hline\hline\hline
Clustering &$CML$ & $RMS$ & $CML$ & $RMS$ \\
\hline
\rule{0pt}{4ex}
Regular& $\widehat{\mathcal{M}}=(0.494,30.190)$ & 4.151 & $\widehat{\mathcal{M}}=(0.491,30.190)$ & 4.153\\
K-means & $\widehat{\mathcal{A}}=(0.001,0.480)$ & & $\widehat{\mathcal{A}}=(0.200,0.473)$ & \\
 & $\widehat{\phi}_1=\left[\begin{array}{cc}
1 & 0\\
0.001 & 0.999
\end{array}\right]$ & & $\widehat{\phi}_1=(0.020,0.980)$ & \\
 & $\widehat{\phi}_2=\left[\begin{array}{ccccc}
1 & 0 & 0 & 0 & 0\\
0.001 & 0.999 & 0 & 0 & 0\\
0.399 & 0.399 & 0.202 & 0 & 0\\
0.299 & 0.300 & 0.299 & 0.102 & 0\\
0.199 & 0.200 & 0.199 & 0.199 & 0.203
\end{array}\right]$ & & $\widehat{\phi}_2=(0.200,0.200,0.201,0.201,0.198)$ & \\
\hline
\rule{0pt}{4ex}
 & $\widehat{\mathcal{M}}=(1.050,13.999)$ & 3.769 &$\widehat{\mathcal{M}}=(1.101,14.099)$ & 3.798 \\
 \envst & $\widehat{\mathcal{A}}=(0.002,0.535)$ & & $\widehat{\mathcal{A}}=(0.199,0.494)$ & \\
& $\widehat{\phi}_1=\left[\begin{array}{cc}
1 & 0\\
0.999 & 0.001
\end{array}\right]$ & & $\widehat{\phi}_1=(0.015,0.985)$ & \\
 & $\widehat{\phi}_2=\left[\begin{array}{ccccc}
1 & 0 & 0 & 0 & 0\\
0.001 & 0.999 & 0 & 0 & 0\\
0.398 & 0.399 & 0.203 & 0 & 0\\
0.299 & 0.299 & 0.299 & 0.103 & 0\\
0.122 & 0.202 & 0.202 & 0.203 & 0.271
\end{array}\right]$ & & $\widehat{\phi}_2=(0.199,0.199,0.204,0.200,0.198)$ & \\
\hline\hline
\end{tabular}.}
\end{center}
\end{table}

\section{Conclusion}

In this article, the new method (\envst) for estimating random environment process $\{z_n\}$ in $RrNGINAR$ $(\mathcal{M,A,P})$ models is defined. The standard K-means clustering method, that was used before, showed poor performances. Taking into account only the values of the data elements, application of the K-means on $RrNGINAR(\mathcal{M,A,P})$ data sequences leads to the loss of  information about random environment. Otherwise, by applying K-means on previously transformed data, the loss of information is si\-gnificantly reduced. This happens because the method follows the behavior of all parameters of the model, which also carry information about belonging to the particular environment state. Hence, \envst\ method leads to a more natural interpretation of the time series, since there is a possibility of finding extre\-mely high or low values in any state. Because of all mentioned above, \envst\ method is more suitable for fine clusterings, where small differences (distances) between means within clusters occur, ie. where  boundaries between states are not straight lines, but wavy or jagged lines. First, the theoretical review of the method is given, with all necessary discussions and clarifications. Appropriate simulated $RrNGINAR(\mathcal{M,A,P})$ time series are created. Application of \envst\ method on the simulated data is implemented. Finally, the supremacy of this new approach over standard K-means is confirmed on popular real-life data.

\section{Acknowledgement} P. Laketa was supported by the OP RDE project “International mobility of research, technical and administrative staff at the Charles University”
CZ.02.2.69/0.0/0.0/18 053/0016976.

\section{Appendix}

\subsection{\textbf{Appendix A}. The choice of model parameters {in the case of }$RrNGINAR(\mathcal{M,A,P})$ simulations with $3$ environment states}

The following parameters combinations are used to create simulated $R3NGINAR(\mathcal{M,A,P})$ time series.
\begin{enumerate}
\item The first combination assumes that means within states are chosen to be close, that is, $\mathcal{M}=(0.5,1,1.5)$. On the other hand, thinning parameters $\alpha_j,\ j=1,2,3,$ differ significantly, with values $\mathcal{A}=(0.1,0.35,0.6)$. Coordinates of the vector $\mathcal{P}=(2,4,2)$ represent maximal orders within states, while corresponding probability matrices are of the form
$$\phi_1=\left[\begin{array}{cc}
1 & 0\\
0.9 & 0.1
\end{array}\right],\ \ \phi_2=\left[\begin{array}{cccc}
1 & 0 & 0 & 0\\
0.2 & 0.8 & 0 & 0\\
0.2 & 0.4 & 0.4 & 0\\
0.2 & 0.2 & 0.3 & 0.3
\end{array}\right],\ \ \phi_3=\left[\begin{array}{cc}
1 & 0\\
0.1 & 0.9
\end{array}\right].$$
Probability matrices given above are used to create \gengmThree\ simulations. To create \gengoThree\  simulations, probabilities located in the last rows of these matrices are going to be exploited. Distribution of the initial state is given as $p_{vec}=(0.3,0.4,0.3)$, while the transition probability matrix favors simulations to remain in the same state, i.e. \[p_{mat}=\left[\begin{array}{ccc}
0.7 & 0.2 & 0.1\\
0.1 & 0.8 & 0.1\\
0.2 & 0.2 & 0.6
\end{array}\right].\] \\


\item The second combination of model parameters will also create an interesting challenge for \envst\ method, since some states have only one pair of parameters which are significantly different. Namely, we have that $\mathcal{M}=(2,4,6)$, $\mathcal{A}=(0.2,0.3,0.6)$ and $\mathcal{P}=(2,4,5)$. Beside that, we have
    $$\phi_1=\left[\begin{array}{cc}
1 & 0\\
0.7 & 0.3
\end{array}\right],\ \ \phi_2=\left[\begin{array}{cccc}
1 & 0 & 0 & 0\\
0.5 & 0.5 & 0 & 0\\
0.3 & 0.3 & 0.4 & 0\\
0.3 & 0.2 & 0.2 & 0.3
\end{array}\right],\ \ \phi_3=\left[\begin{array}{ccccc}
1 & 0 & 0 & 0 & 0\\
0.4 & 0.6 & 0 & 0 & 0\\
0.2 & 0.5 & 0.3 & 0 & 0\\
0.25 & 0.3 & 0.2 & 0.25 & 0\\
0.2 & 0.2 & 0.3 & 0.1 & 0.2
\end{array}\right].$$
As we can see, means within states grow progressively, although the jumps are not too high. The first and the second state have similar thinning parameters, while corresponding orders differ significantly. On the other hand, the second and the third state have similar orders, while corresponding thinning parameters differ significantly. Finally, all parameters of the first and the third state differ significantly. In order to be able to place the realization at moment $n$ in the appropriate cluster, it is crucial for the clustering method to possess information about the behavior of all parameters of the model at the same moment.\\
An initial state has the distribution $p_{vec}=(0.35,0.35,0.3)$ and the transition probability matrix is of the form \[p_{mat}=\left[\begin{array}{ccc}
0.9 & 0.05 & 0.05\\
0.2 & 0.7 & 0.1\\
0.1 & 0.1 & 0.8
\end{array}\right].\]

\end{enumerate}

\subsection{\textbf{Appendix B}. Simulation study {in the case of }simulated $RrNGINAR(\mathcal{M,A,P})$ time series with $3$ environment states}

\begin{figure}
\centering
\begin{minipage}[b]{0.85\textwidth}
	 	\includegraphics[width=1\linewidth]{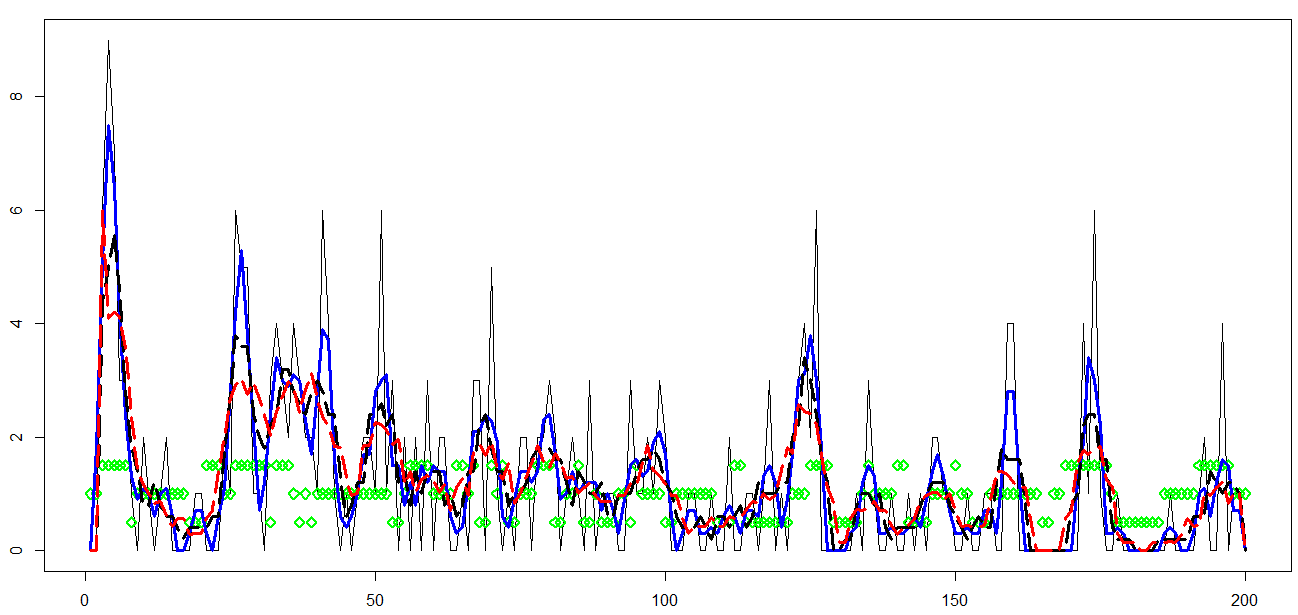}
	         	\caption*{\footnotesize a) $R3NGINAR_{max}(2,4,2)$ model}
 \end{minipage}
\begin{minipage}[b]{0.85\textwidth}
	 	\includegraphics[width=1\linewidth]{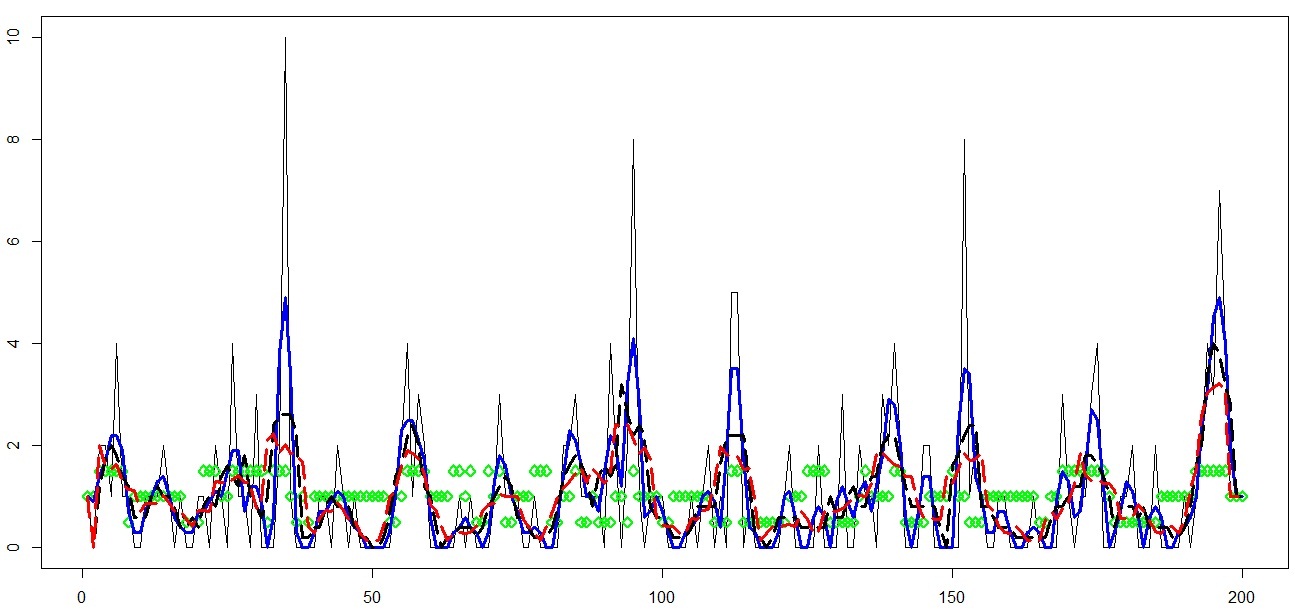}
	         	\caption*{\footnotesize b) $R3NGINAR_1(2,4,2)$ model}
 \end{minipage}
\caption{\footnotesize Pre-estimates of $\{{\mu_n}\}$ obtained for various selections of $\mathbf{c}_m$ in the case of simulated $R3NGINAR(2,4,2)$ models: green diamond-exact mean values sequence $\{\mu_n\}$; regular black line-sequence $\{T(\tilde{\mu}_n,\mathbf{c}_m)\}$ for $\mathbf{c}_m=1$; thick blue line- sequence $\{T(\tilde{\mu}_n,\mathbf{c}_m)\}$ for $\mathbf{c}_m=(0.4,0.3)$; dashed black line-sequence $\{T(\tilde{\mu}_n,\mathbf{c}_m)\}$ for $\mathbf{c}_m=(0.2,0.2,0.2)$, dashed red line-sequence $\{T(\tilde{\mu}_n,\mathbf{c}_m)\}$ for $\mathbf{c}_m=(0.16,0.14,0.14,0.14)$.}
\label{fig:first comb1}
\end{figure}

Simulation study follows the same path as it was the case with testing on simulated data with $2$ environment states.  After creating $R3NGINAR_{max}(2,4,2)$ and $R3NGINAR_{1}(2,4,2)$  simulations (two replications of each) using the first combination of model parameters, one may start with determination of \envst\ method parameters. The sequence  $\{\tilde{\mu}_n\}$ is again obtained by usage of (\ref{mu}). To improve such obtained sequence of pre-estimates, optimal shape of the vector $\mathbf{c}_m$ is of interest. Sequences $\{T(\tilde{\mu}_n,\mathbf{c}_m)\}$ obtained for various selections of $\mathbf{c}_m$ are shown in Figure \ref{fig:first comb1} and compared to the real parameter values $\{\mu_n\}$. \\

For both simulations, $R3NGINAR_{max}(2,4,2)$ and $R3NGINAR_{1}(2,4,2)$, the best pre-estimate result is obtained {in the case of }$\mathbf{c}_m=(0.16,0.14,0.14,0.14)$. The ability of trimming the high peaks is noticed. In particular, this pre-estimate shows remarkable potential to assess means within the second (middle) state.\\

Further, a determination of the sequence $\{\tilde{P}_n\}$ takes place in two steps. The first step towards that goal is to determine the value of parameter $d_p$. Calculations of $\Delta_p$ for various selections of $d_p$ are performed and correspo\-nding results are given in Table \ref{tab}. Optimal $d_p$ values are $17$ and $18$ (for $R3NGINAR_{max}(2,4,2)$ and $R3NGINAR_{1}(2,4,2)$  respectively). \\

\begin{table}[htbp]
{\caption{\footnotesize Values of the error $\Delta_p$ for various selections of $d_p$}\label{tab}{
{\scriptsize\vskip 3mm\begin{center}  \begin{tabular}{|c|c|c|c|c|c|c|c|}\hline

 \multicolumn{2}{c}{ $R3NGINAR_{max}(2,4,2)$   }\vline & \multicolumn{2}{c}{$R3NGINAR_{1}(2,4,2)$} \vline& \multicolumn{2}{c}{ $R3NGINAR_{max}(2,4,5)$   }\vline & \multicolumn{2}{c}{ $R3NGINAR_{1}(2,4,5)$    } \\
\hline
 $d_p$ & $\Delta_p$ &  $d_p$ & $\Delta_p$ & $d_p$ & $\Delta_p$ &  $d_p$ & $\Delta_p$ \\
\hline
  5 & 1.571 & 5 & 1.605 & 5 & 1.925 & 5 & 1.881  \\
  6 & 1.569 & 6 & 1.695 & 6 & 1.915 & 6 & 1.906 \\
  7 & 1.573 & 7 & 1.582 & 7 & 1.912 & 7 & 1.918 \\
  8 & 1.523 & 8 & 1.580 & 8 & 1.882 & 8 & 1.932 \\
  9 & 1.568 & 9 & 1.569 & 9 & 1.823 & 9 & 1.887 \\
  10 & 1.556 & 10 & 1.596 & 10 & 1.785 & 10 & 1.861 \\
  11 & 1.520 & 11 & 1.565 & 11 & 1.749 & \textbf{11} & \textbf{1.850} \\
  12 & 1.503 & 12 & 1.572 & \textbf{12} & \textbf{1.731} & 12 & 1.890 \\
  13 & 1.504 & 13 & 1.536 & 13 & 1.774 & 13 & 1.881 \\
  14 & 1.464 & 14 & 1.474 & 14 & 1.807 & 14 & 1.905 \\
  15 & 1.425 & 15 & 1.474 & 15 & 1.798 & 15 & 1.911 \\
  16 & 1.430 & 16 & 1.456 & 16 & 1.810 & 16 & 1.916 \\
  \textbf{17} & \textbf{1.416} & 17 & 1.474 & 17 & 1.823 & 17 & 1.950 \\
  18 & 1.418 & \textbf{18} & \textbf{1.408} & 18 & 1.844 & 18 & 1.944\\
  19 & 1.420 & 19 & 1.455 & 19 & 1.854 & 19 & 1.916 \\
  20 & 1.418 & 20 & 1.433 & 20 & 1.875 & 20 & 1.884 \\
\hline
\end{tabular}
\end{center}}}}
\end{table}

\begin{figure}
\centering
\begin{minipage}[b]{0.85\textwidth}
	 	\includegraphics[width=1\linewidth]{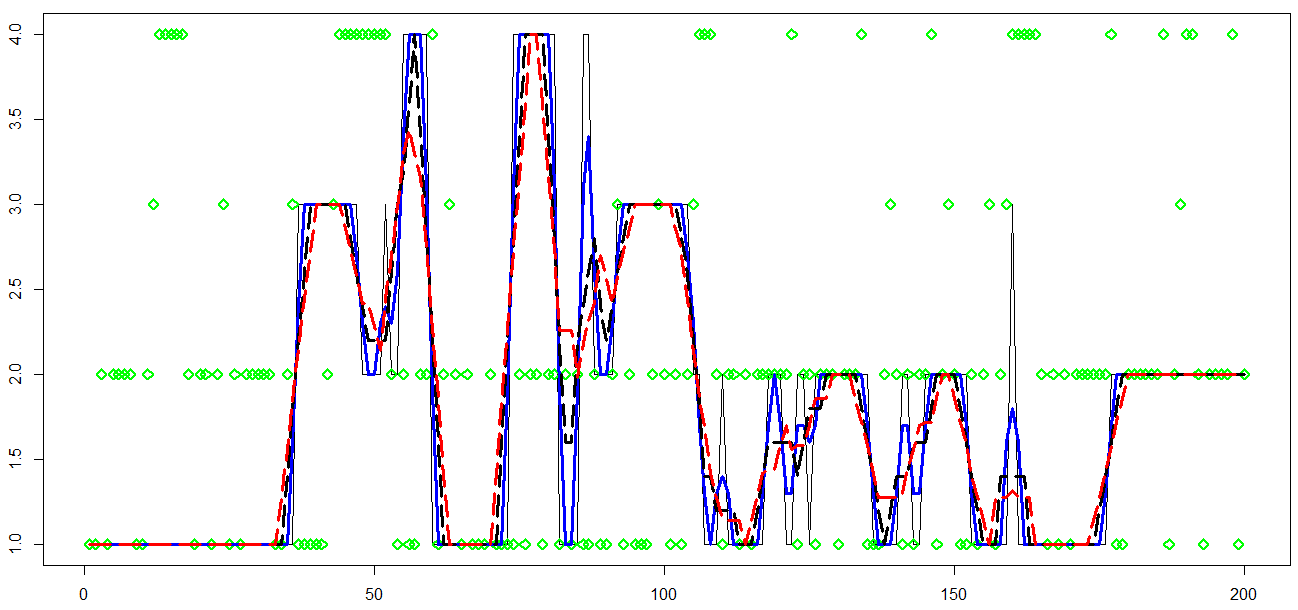}
	         	\caption*{\footnotesize a) $R3NGINAR_{max}(2,4,2)$ model with $d_p=17$}
 		
 \end{minipage}
\begin{minipage}[b]{0.85\textwidth}
	 	\includegraphics[width=1\linewidth]{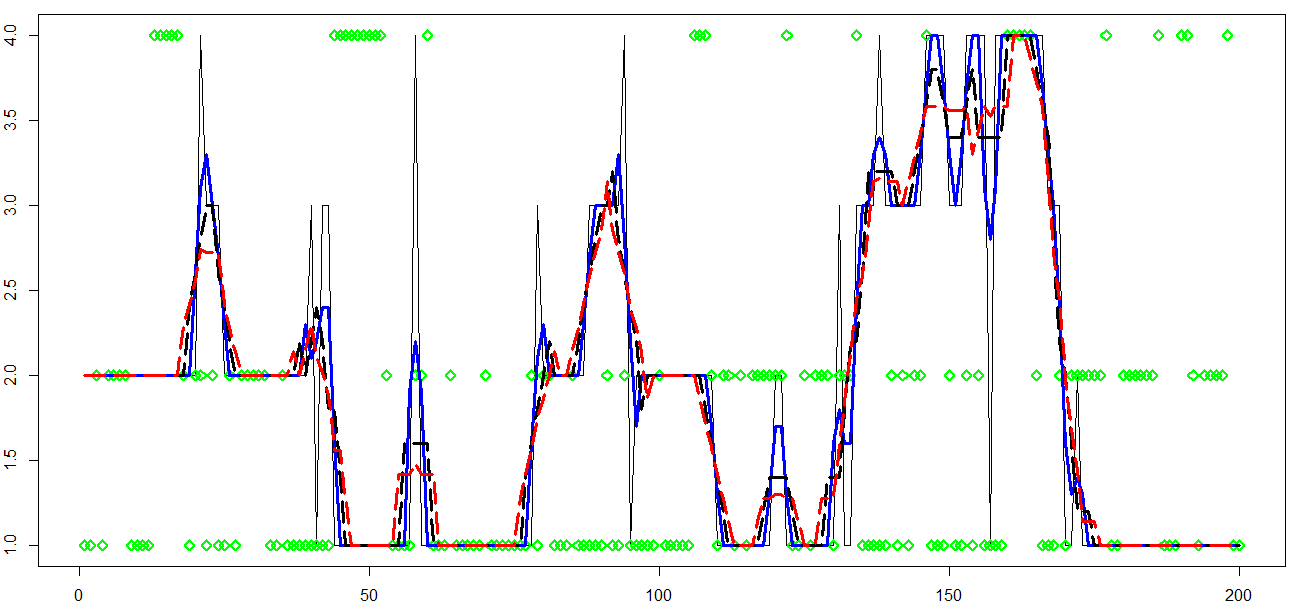}
	         	\caption*{\footnotesize b) $R3NGINAR_{1}(2,4,2)$ model with $d_p=18$}
 		\end{minipage}
		\caption{\footnotesize Pre-estimates of $\{P_n\}$ obtained for various selections of $\mathbf{c}_p$ in the case of simulated $R3NGINAR(2,4,2)$ models: green diamond-exact order sequence $\{P_n\}$; regular black line-sequence $\{T(\tilde{P}_n,\mathbf{c}_p)\}$ for $\mathbf{c}_p=1$; thick blue line-sequence $\{T(\tilde{P}_n,\mathbf{c}_p)\}$ for $\mathbf{c}_p=(0.4,0.3)$; dashed black line-sequence $\{T(\tilde{P}_n,\mathbf{c}_p)\}$ for $\mathbf{c}_p=(0.2,0.2,0.2)$, dashed red line-sequence $\{T(\tilde{P}_n,\mathbf{c}_p)\}$ for $\mathbf{c}_p=(0.16,0.14,0.14,0.14)$.}
		\label{fig:cp11}
\end{figure}

The second step in determination of $\{\tilde{P}_n\}$ is to provide corresponding vector $\mathbf{c}_p$ for fixed optimal value of $d_p$. Sequences $\{T(\tilde{P}_n,\mathbf{c}_p)\}$ obtained for various selections of $\mathbf{c}_p$ are shown in Figure~\ref{fig:cp11} and compared to the exact order sequence $\{P_n\}$. According to figure, sequences obtained in case when $k=2$, $k=3$ and $k=4$ behave practically the same. On the other hand, the sequence obtained for $k=1$ has frequent and sharp ups and downs, which lead to the erroneous clustering result. The same conclusion holds for both simulations, $R3NGINAR_{max}(2,4,2)$ and $R3NGINAR_{1}(2,4,2)$. Due to the simplicity of the model, we prefer to take $k=2$ and $\mathbf{c}_p=(0.4,0.3)$. \\

\begin{figure}
\centering
\begin{minipage}[b]{0.85\textwidth}
	 	\includegraphics[width=1\linewidth]{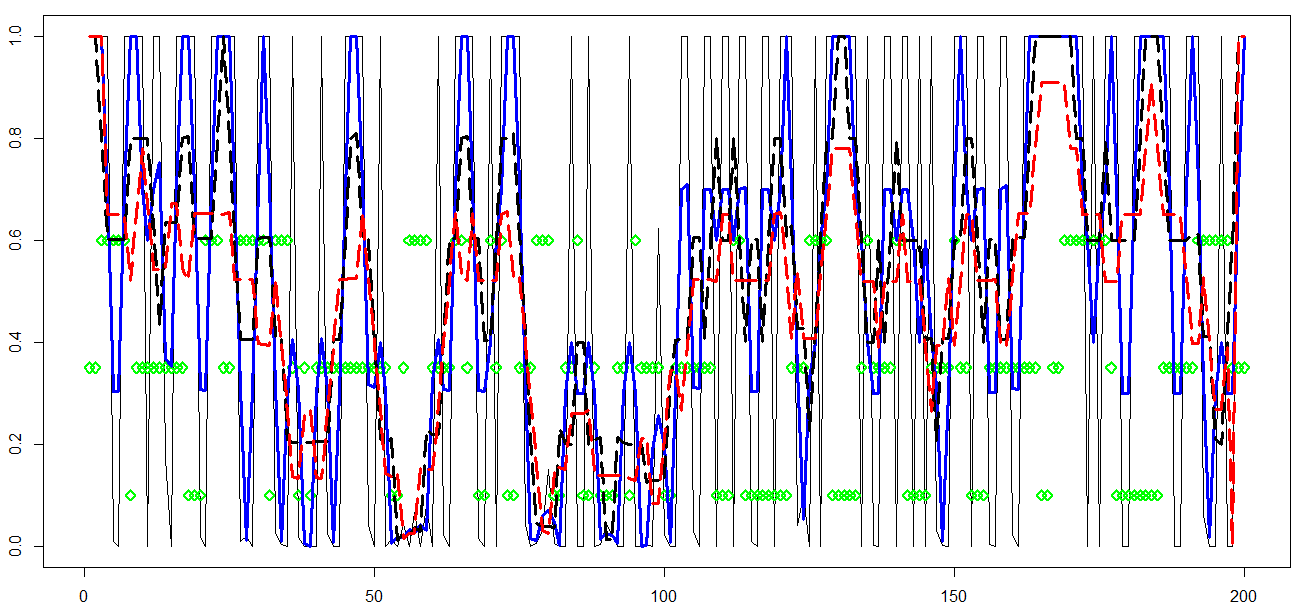}
	         	\caption*{\footnotesize a) $R3NGINAR_{max}(2,4,2)$ model}
 		\label{slika5}
 \end{minipage}
\begin{minipage}[b]{0.85\textwidth}
	 	\includegraphics[width=1\linewidth]{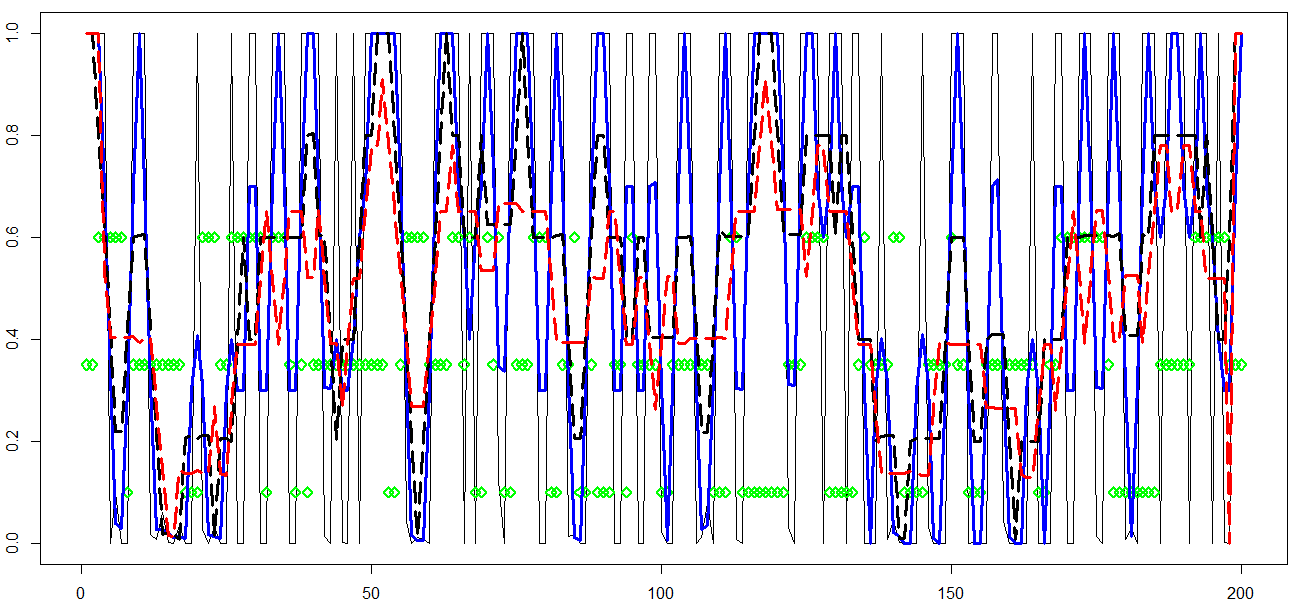}
	         	\caption*{\footnotesize b) $R3NGINAR_{1}(2,4,2)$ model}
 		\label{slika6}
	\end{minipage}
	\caption{\footnotesize Pre-estimates of $\{\alpha_n\}$ obtained for various selections of $\mathbf{c}_a$ in the case of simulated $R3NGINAR(2,4,2)$ models: green diamond-exact thinning parameters sequence $\{\alpha_n\}$; regular black line-sequence $\{T(\tilde{\alpha}_n,\mathbf{c}_a)\}$ for $\mathbf{c}_a=1$; thick blue line-sequence $\{T(\tilde{\alpha}_n,\mathbf{c}_a)\}$ for $\mathbf{c}_a=(0.4,0.3)$; dashed black line-sequence $\{T(\tilde{\alpha}_n,\mathbf{c}_a)\}$ for $\mathbf{c}_a=(0.2,0.2,0.2)$, dashed red line-sequence $\{T(\tilde{\alpha}_n,\mathbf{c}_a)\}$ for $\mathbf{c}_a=(0.16,0.14,0.14,0.14)$.}
	\label{fig:ca111}
 \end{figure}

We are now able to determine $\tilde{\alpha}_n,\ n\in N,$ using (\ref{alpha}). To improve such obtained pre-estimates, determination of $\mathbf{c}_a$ took place. Sequences $\{T(\tilde{\alpha}_n,\mathbf{c}_a)\}$ obtained for various selections of $\mathbf{c}_a$ are given in Figure~\ref{fig:ca111} and compared to the real sequence $\{\alpha_n\}$. As figure shows, sequence of pre-estimates obtained for $k = 4$ is in advantage in regard to other sequences. For $\mathbf{c}_a=(0.16,0.14,0.14,0.14)$, just a few steep jumps are located on the plot curve. Pre-estimates are rarely beyond the greatest thinning parameter value, and even if something like that happens, the overdrafts are generally not large. Most of the time, this sequence of pre-estimates keeps oscillating between $\alpha_1$ and $\alpha_3$, with particulary god assessment of $\alpha_2$. The same conclusion holds for both, $R3NGINAR_{max}(2,4,2)$ and $R3NGINAR_{1}(2,4,2)$ simulation. Thus, in both cases we have $\mathbf{c}_a=(0.16,0.14,0.14,0.14)$. \\

\begin{figure}
\centering
\begin{minipage}[b]{0.69\textwidth}
	 	\includegraphics[width=1\linewidth]{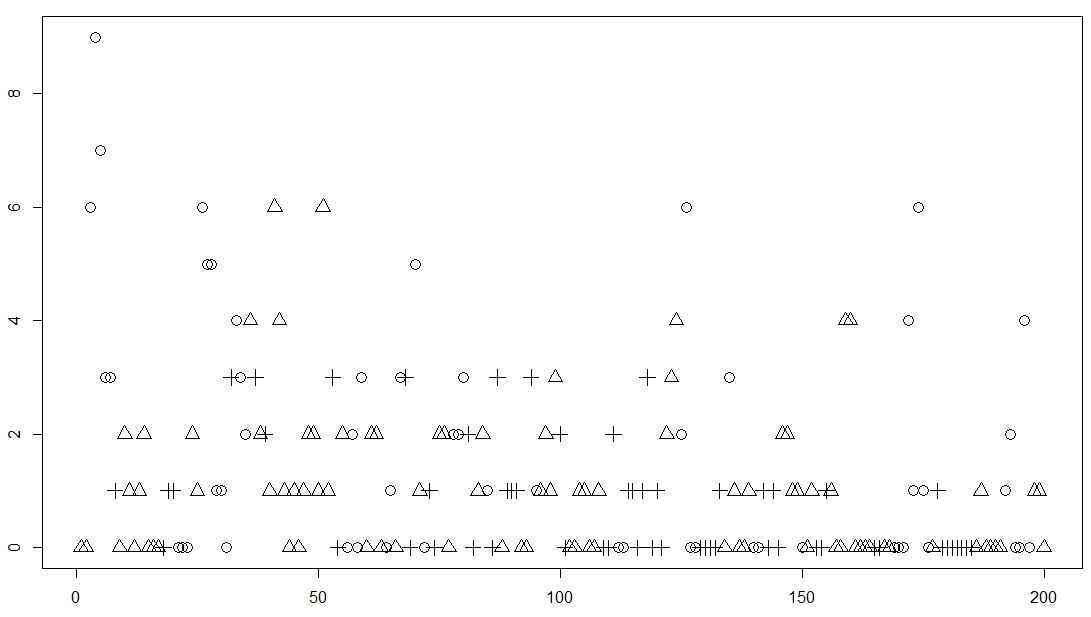}
	         	\caption*{\footnotesize Exact states of $R3NGINAR_{max}(2,4,2)$ simulation}
 \end{minipage}
\begin{minipage}[b]{0.69\textwidth}
	 	\includegraphics[width=1\linewidth]{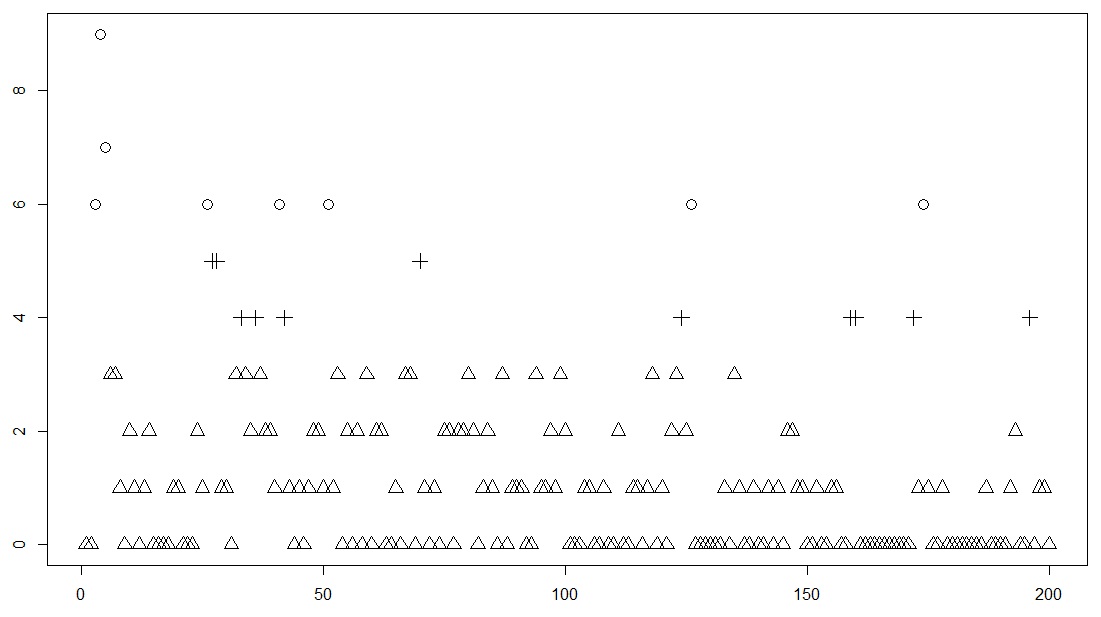}
	         	\caption*{\footnotesize States obtained by standard K-means clustering method}
 \end{minipage}
\begin{minipage}[b]{0.69\textwidth}
	 	\includegraphics[width=1\linewidth]{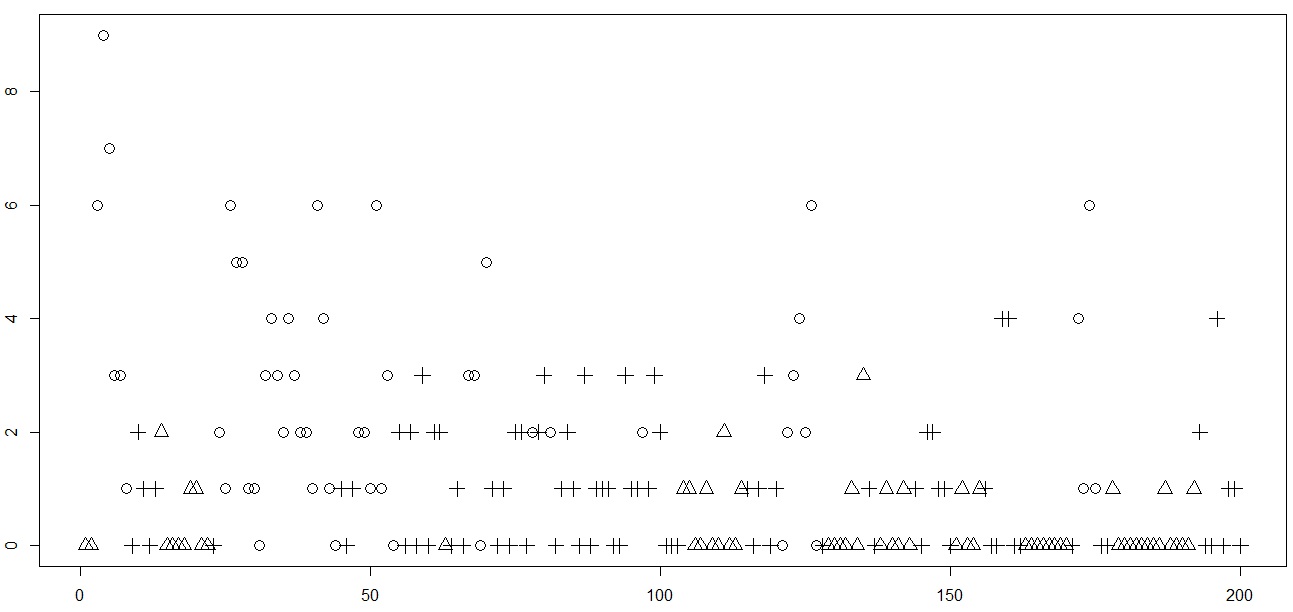}
	         	\caption*{\footnotesize States obtained by \envst\ method for $d_p=17$, $\mathbf{c}_m=(0.16,0.14,0.14,0.14)$, $\mathbf{c}_a=(0.16,0.14,0.14,0.14)$, $\mathbf{c}_p=(0.4,0.3)$, $C_m=9$, $C_a=7$, $C_p=2$}
 \end{minipage}
\caption{\footnotesize The environment states of $R3NGINAR_{max}(2,4,2)$ simulation}
\label{fig:renevMax1}
\end{figure}

\begin{figure}
\centering
\begin{minipage}[b]{0.69\textwidth}
	 	\includegraphics[width=1\linewidth]{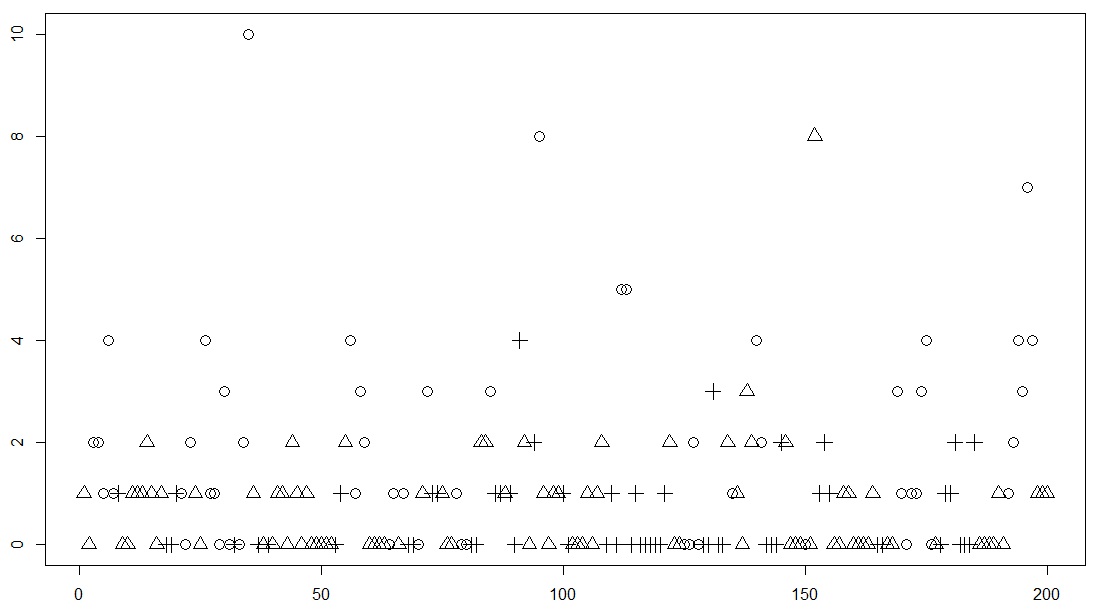}
	         	\caption*{\footnotesize Exact states of $R3NGINAR_{1}(2,4,2)$ simulation}
\end{minipage}
\begin{minipage}[b]{0.69\textwidth}
	 	\includegraphics[width=1\linewidth]{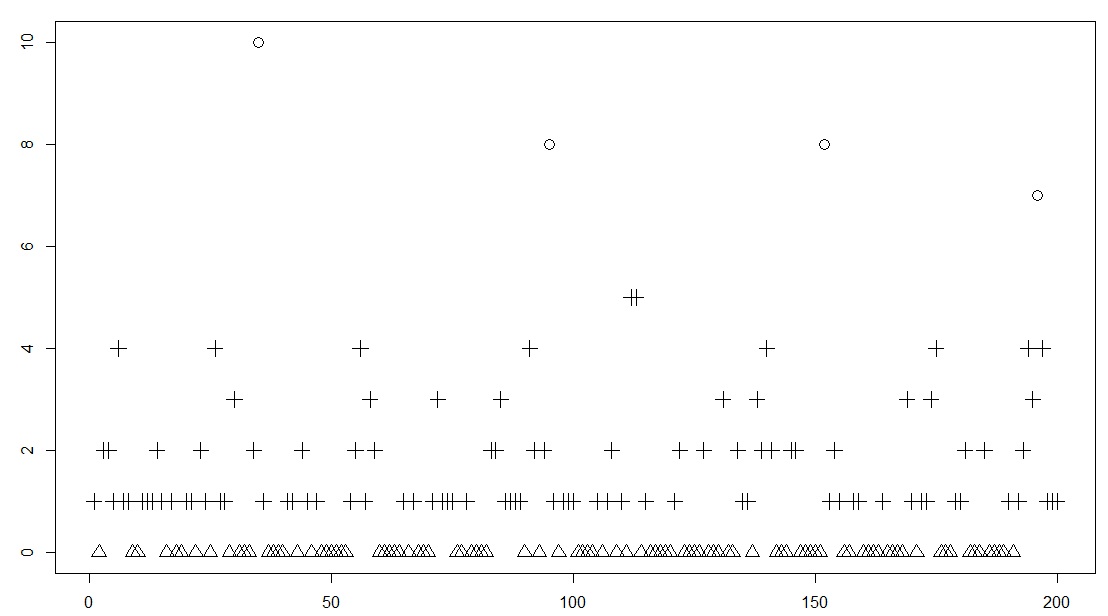}
	         	\caption*{\footnotesize States obtained by standard K-means clustering method}
  \end{minipage}
\begin{minipage}[b]{0.69\textwidth}
	 	\includegraphics[width=1\linewidth]{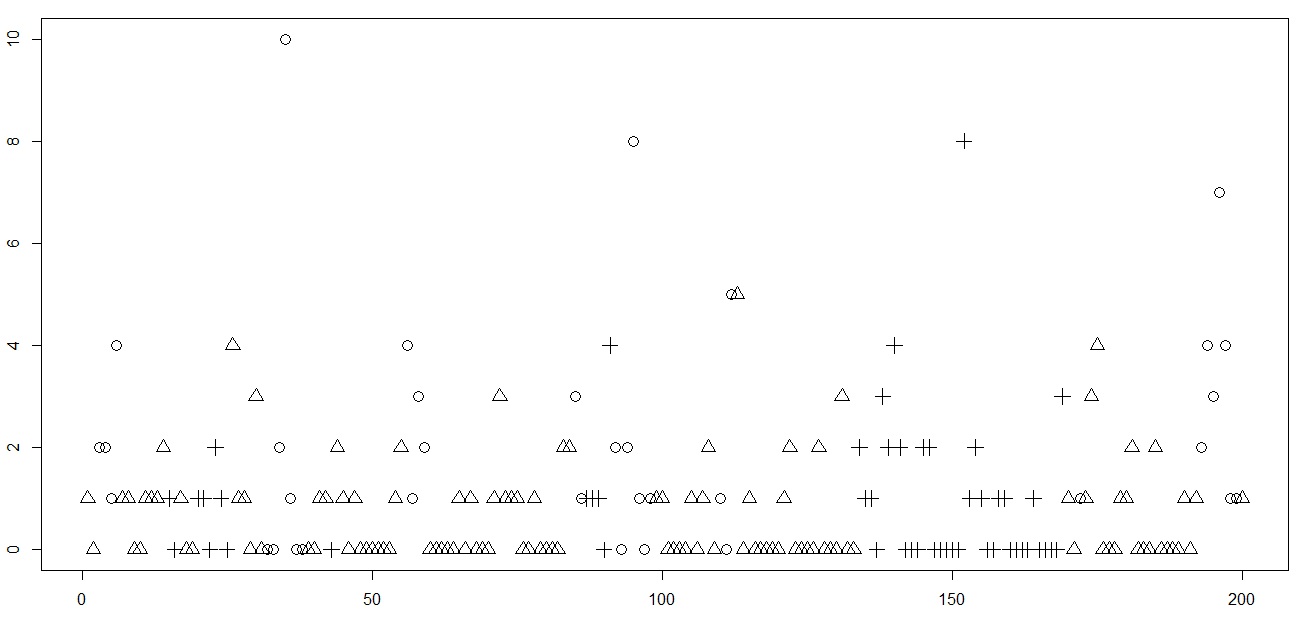}
	         	\caption*{\footnotesize States obtained by \envst\ method for $d_p=18$, $\mathbf{c}_m=(0.16,0.14,0.14,0.14)$, $\mathbf{c}_a=(0.16,0.14,0.14,0.14)$, $\mathbf{c}_p=(0.4,0.3)$, $C_m=6$, $C_a=1$, $C_p=8$}
\end{minipage}
\caption{\footnotesize The environment states of $R3NGINAR_{1}(2,4,2)$ simulation}
\label{fig:renev111}
\end{figure}

Previous results regarding $d_p$, $\mathbf{c}_m$, $\mathbf{c}_a$ and $\mathbf{c}_p$ are summarized in Table \ref{pomocna tabela3}. To determine $C_m$, $C_a$ and $C_p$, the clustering of
\begin{equation}\nonumber
\{(C_mS(\tilde{\mu}_n,\mathbf{c}_m),C_aS(\tilde{\alpha}_n,\mathbf{c}_a),C_pS(\tilde{P}_n,\mathbf{c}_p))\}
\end{equation}
is performed for each $C_m=i$, $C_a=j$, $C_p=l$, $i,j,l=1,2,\ldots, 10,$ and thousand different estimates of $\{z_n\}$ are provided. The best result is obtained for $C_m=9$, $C_a=7$ and $C_p=2$ {in the case of }simulated $R3NGINAR_{max}(2,4,2)$ time series, having $209$ estimated states which are equal to corresponding exact states. On contrary to that, the standard K-means method managed to have only $155$ exactly estimated states, which doesn't seem acceptable at all. A comparative overview of exact states, states obtained by standard K-means method and states obtained by usage of new \envst\ method is provided by Figure \ref{fig:renevMax1}. The same procedure is applied {in the case of }simulated $R3NGINAR_{1}(2,4,2)$  time series. The highest number of exactly estimated states is obtained for $C_m=6$, $C_a=1$ and $C_p=8$, with $216$ elements which estimated states are equal to corresponding exact states. The standard K-means method managed to have only $153$ exactly estimated states. A comparative overview of exact states, states obtained by standard K-means method and states obtained by usage of \envst\ method is provided by Figure \ref{fig:renev111}.\\

\begin{table}[htbp]
\caption{\footnotesize Values of the constant $d_p$ and vectors $\mathbf{c}_m$, $\mathbf{c}_a$, $\mathbf{c}_p$, {in the case of }simulated $R3NGINAR(2,4,2)$ time series}
\begin{center}\label{pomocna tabela3}{\footnotesize\begin{tabular}{|c|c|c|c|}
\hline
\multicolumn{4}{c}{$R3NGINAR_{max}(2,4,2)$}\\
\hline
$d_p$& $\mathbf{c}_m$&$\mathbf{c}_a$& $\mathbf{c}_p$\\
\hline
17&(0.16,0.14,0.14,0.14) & (0.16,0.14,0.14,0.14)& (0.4,0.3)\\
\hline
\multicolumn{4}{c}{$R3NGINAR_1(2,4,2)$ }\\
\hline
$d_p$& $\mathbf{c}_m$&$\mathbf{c}_a$& $\mathbf{c}_p$\\
\hline
18&(0.16,0.14,0.14,0.14) & (0.16,0.14,0.14,0.14)& (0.4,0.3)\\
\hline
\end{tabular}.}
\end{center}
\end{table}

Same as it was the case with $2$ environment states simulations, several improvements are noticeable here as well. Beside higher number of exactly estimated states, the \envst\ method produces much longer sequences of consecutive elements in each of $3$ given states. In general, this improvement enables more successful application of random environment $INAR$ models of higher order. Further, the possibility of finding extremely high or low values in any of three given states is perceptible here. Furthermore, the possibility of having equal elements in different states is also detected. Since this often occurs in generalized random environment \inar\ time series of higher order with similar means within states, the \envst\ method seems more applicable than the standard K-means.\\

The amount of benefit one gets by applying the \envst\ method is measured on unused replications of the simulated $R3NGINAR_{max}(2,4,2)$ and $R3NGINAR_{1}(2,4,2)$ time series. Having results of both environment state estimation methods mentioned earlier, a valid reconstruction of given simulations is performed by appropriate $R3NGINAR_{max}(2,4,2)$ or $R3NGINAR_{1}(2,4,2)$ model for each clustering result singularly. $RMS$ of differences between simulated data and their reconstructions should indicate whether there is any truly benefit from applying the \envst\ method. Results of modeling obtained after applying the standard K-means method and after applying the \envst\ method are provided in Table \ref{tabela4}. Although a way smaller than {in the case of }simulations with $2$ environment states, the benefit of applying the \envst\ method still exists. Reconstructions after K-means clu\-stering produced the following $RMS$ values: $RMS=1.285$ {in the case of }$R3NGINAR_{max}(2,4,2)$ simulation and $RMS=1.471$ {in the case of }$R3NGINAR_{1}(2,4,2)$  simulation. On contrary to that, reconstructions after \envst\ method produced the following: $RMS=1.149$ {in the case of }$R3NGINAR_{max}(2,4,2)$ simulation and $RMS=1.370$ {in the case of }$R3NGINAR_{1}(2,4,2)$  simulation.\\ 

\begin{table}[htbp]
\caption{\footnotesize $CML$ parameter estimates and $RMS$ values obtained after reconstruction of the simulated data sequences that correspond to the $R3NGINAR_{max}(2,4,2)$ and $R3NGINAR_{1}(2,4,2)$ time series}
\begin{center}\label{tabela4}{\footnotesize\begin{tabular}{|c|c|c|c|c|}
\hline
 & \multicolumn{2}{c}{ $R3NGINAR_{max}(2,4,2)$   }\vline &\multicolumn{2}{c}{ $R3NGINAR_1(2,4,2)$ }\vline\\
\hline\hline
Clustering &$CML$ & $RMS$ & $CML$ & $RMS$ \\
\hline\hline
\rule{0pt}{4ex}
Regular& $\widehat{\mathcal{M}}=(0.324,2.308,5.278)$ & 1.285 & $\widehat{\mathcal{M}}=(0.523,3.558,7.779)$ & 1.471\\
K-means & $\widehat{\mathcal{A}}=(0.051,0.188,0.2199)$ & & $\widehat{\mathcal{A}}=(0.052,0.201,0.437)$ & \\
 & $\widehat{\phi}_1=\left[\begin{array}{cc}
1 & 0\\
0.894 & 0.106
\end{array}\right]$ & & $\widehat{\phi}_1=(0.999,0.001)$ & \\
 & $\widehat{\phi}_2=\left[\begin{array}{cccc}
1 & 0 & 0 & 0\\
0.001 & 0.999 & 0 & 0\\
0.331 & 0.331 & 0.338 & 0\\
0.220 & 0.202 & 0.240 & 0.338
\end{array}\right]$ & & $\widehat{\phi}_2=(0.249,0.201,0.241,0.309)$ & \\
 & $\widehat{\phi}_3=\left[\begin{array}{cc}
1 & 0\\
0.001 & 0.999
\end{array}\right]$ & & $\widehat{\phi}_3=(0.001,0.999)$ & \\
\hline
\rule{0pt}{4ex}
 & $\widehat{\mathcal{M}}=(0.501,1.201,1.503)$ & 1.149 &$\widehat{\mathcal{M}}=(0.503,1.161,1.449)$ & 1.370 \\
 \envst\ & $\widehat{\mathcal{A}}=(0.199,0.328,0.321)$ & & $\widehat{\mathcal{A}}=(0.090,0.213,0.368)$ & \\
& $\widehat{\phi}_1=\left[\begin{array}{cc}
1 & 0\\
0.983 & 0.017
\end{array}\right]$ & & $\widehat{\phi}_1=(0.754,0.246)$ & \\
 & $\widehat{\phi}_2=\left[\begin{array}{cccc}
1 & 0 & 0 & 0\\
0.001 & 0.999 & 0 & 0\\
0.326 & 0.331 & 0.343 & 0\\
0.243 & 0.201 & 0.242 & 0.314
\end{array}\right]$ & & $\widehat{\phi}_2=(0.256,0.211,0.247,0.286)$ & \\
 & $\widehat{\phi}_3=\left[\begin{array}{cc}
1 & 0\\
0.001 & 0.999
\end{array}\right]$ & & $\widehat{\phi}_3=(0.141,0.859)$ & \\
\hline\hline
\end{tabular}}.
\end{center}
\end{table}

In order to confirm additionally the effectiveness of the \envst\ method, two replications of the simulated $R3NGINAR_{max}(2,4,5)$ and $R3NGINAR_{1}(2,4,5)$ time series were created, dictated by the second combination of model parameters. Optimal values of $d_p$, $\mathbf{c}_m$, $\mathbf{c}_a$ $\mathbf{c}_p$, $C_m$, $C_a$ and $C_p$ were obtained, based on the first replications of each pair. The procedure presented in previous cases was followed this time as well. Results thus obtained are presented in Table \ref{pomocna tabela4}.\\

\begin{table}[htbp]
\caption{\footnotesize Values of the constant $d_p$ and vectors $\mathbf{c}_m$, $\mathbf{c}_a$, $\mathbf{c}_p$, {in the case of }simulated $R3NGINAR(2,4,5)$ time series }
\begin{center}\label{pomocna tabela4}{\footnotesize\begin{tabular}{|c|c|c|c|c|c|c|}
\hline
\multicolumn{7}{c}{$R3NGINAR_{max}(2,4,5)$}\\
\hline
$d_p$& $\mathbf{c}_m$&$\mathbf{c}_a$& $\mathbf{c}_p$ & $C_m$ & $C_a$ & $C_p$\\
\hline
12&(0.16,0.14,0.14,0.14) & (0.16,0.14,0.14,0.14)& (0.4,0.3) & 10 & 3 & 1\\
\hline
\multicolumn{7}{c}{$R3NGINAR_{1}(2,4,5)$ }\\
\hline
$d_p$& $\mathbf{c}_m$&$\mathbf{c}_a$& $\mathbf{c}_p$ & $C_m$ & $C_a$ & $C_p$\\
\hline
11&(0.16,0.14,0.14,0.14) & (0.16,0.14,0.14,0.14)& (0.4,0.3) & 7 & 5 & 2\\
\hline
\end{tabular}.}
\end{center}
\end{table}

Further, the standard K-means and the \envst\  method are performed on unused replications. Furthermore, those replications are reconstructed by corresponding $R3NGINAR_{max}(2,4,5)$ or $R3NGINAR_{1}(2,4,5)$ model for each clustering result, and modeling results such obtained are provided in Table \ref{tabela41}. As one may notice, $RMS$ values are not that large in general, bearing on mind relatively high realization values. Hence, the amount of benefit detected after application of the \envst\ method is really satisfactory. The benefits are mainly generated by a more accurate estimates of the mean values. As for the other model parameters, corresponding estimates are of the same level.

\begin{table}[htbp]
\caption{\footnotesize $CML$ parameter estimates and $RMS$ values obtained after reconstruction of the simulated data sequences that correspond to the $R3NGINAR_{max}(2,4,5)$ and $R3NGINAR_{1}(2,4,5)$ time series}
\begin{center}\label{tabela41}{\footnotesize\begin{tabular}{|c|c|c|c|c|}
\hline
 & \multicolumn{2}{c}{ $R3NGINAR_{max}(2,4,5)$   }\vline &\multicolumn{2}{c}{ $R3NGINAR_1(2,4,5)$ }\vline\\
\hline\hline
Clustering &$CML$ & $RMS$ & $CML$ & $RMS$ \\
\hline\hline
\rule{0pt}{4ex}
Regular& $\widehat{\mathcal{M}}=(0.760,4.420,11.410)$ & 2.109 & $\widehat{\mathcal{M}}=(0.770,5.060,14.625)$ & 2.353\\
K-means & $\widehat{\mathcal{A}}=(0.177,0.300,0.299)$ & & $\widehat{\mathcal{A}}=(0.164,0.200,0.302)$ & \\
 & $\widehat{\phi}_1=\left[\begin{array}{cc}
1 & 0\\
0.008 & 0.992
\end{array}\right]$ & & $\widehat{\phi}_1=(0.009,0.991)$ & \\
 & $\widehat{\phi}_2=\left[\begin{array}{cccc}
1 & 0 & 0 & 0\\
0.968 & 0.032 & 0 & 0\\
0.299 & 0.300 & 0.401 & 0\\
0.299 & 0.300 & 0.300 & 0.101
\end{array}\right]$ & & $\widehat{\phi}_2=(0.301,0.302,0.300,0.097)$ & \\
 & $\widehat{\phi}_3=\left[\begin{array}{ccccc}
1 & 0 & 0 & 0 & 0\\
0.001 & 0.999 & 0 & 0 & 0\\
0.399 & 0.399 & 0.202 & 0 & 0\\
0.299 & 0.300 & 0.299 & 0.102 & 0\\
0.199 & 0.199 & 0.200 & 0.201 & 0.201
\end{array}\right]$ & & $\widehat{\phi}_3=(0.202,0.200,0.201,0.202,0.195)$ & \\
\hline
\rule{0pt}{4ex}
 & $\widehat{\mathcal{M}}=(2.496,4.499,6.511)$ & 1.719 &$\widehat{\mathcal{M}}=(2.369,4.426,6.495)$ & 1.977 \\
 \envst\ & $\widehat{\mathcal{A}}=(0.298,0.300,0.424)$ & & $\widehat{\mathcal{A}}=(0.162,0.198,0.473)$ & \\
& $\widehat{\phi}_1=\left[\begin{array}{cc}
1 & 0\\
0.001 & 0.999
\end{array}\right]$ & & $\widehat{\phi}_1=(0.449,0.551)$ & \\
 & $\widehat{\phi}_2=\left[\begin{array}{cccc}
1 & 0 & 0 & 0\\
0.477 & 0.523 & 0 & 0\\
0.298 & 0.299 & 0.403 & 0\\
0.298 & 0.300 & 0.299 & 0.103
\end{array}\right]$ & & $\widehat{\phi}_2=(0.294,0.198,0.259,0.249)$ & \\
 & $\widehat{\phi}_3=\left[\begin{array}{ccccc}
1 & 0 & 0 & 0 & 0\\
0.001 & 0.999 & 0 & 0 & 0\\
0.399 & 0.400 & 0.201 & 0 & 0\\
0.299 & 0.300 & 0.300 & 0.101 & 0 \\
0.199 & 0.199 & 0.200 & 0.201 & 0.201
\end{array}\right]$ & & $\widehat{\phi}_3=(0.239,0.162, 0.250, 0.161,0.186)$ & \\
\hline\hline
\end{tabular}}.
\end{center}
\end{table}

\subsection{\textbf{Appendix C}. Application of various models with stationary or non-stationary nature }

Beside mentioned $R2NGINAR(2,4)$ and $R2NGINAR(2,5)$ models, several models with stationary or non-stationary nature are considered here. The following stationary models are taken into account: $INAR(1)$ model with Poisson marginals ($PoINAR(1)$) from \cite{AlAlzaid}, quasi-binomial $INAR(1)$ model with generalized Poisson marginals ($GPQINAR(1)$) presented in \cite{AA1993}, geometric $INAR(1)$ model ($GINAR(1)$) provided by \cite{AlAlOsh}, new geometric $INAR(1)$ model ($NGINAR(1)$) defined by \cite{RBN2009}, combined geometric $INAR(p)$ model ($NGINAR(p)$) given in \cite{NRB2012}, $p=2,3,4,5,$  and random coefficient $INAR(1)$ model with negative binomial marginals ($NBRCINAR(1)$) defined by \cite{ZengBasawaData2007}. As for the non-stationary models, we consider the following: a $2$ state random environment $NGINAR(1)$ model presented in \cite{NLR2016} and random environment models of higher order ($R2NGINAR(p)$) described in \cite{NLR2017}, where $p=2,3,4,5$. \\

Corresponding modeling results are given in Table \ref{tabRealniPod} and Table \ref{tabela221}. Table \ref{tabRealniPod} contains results obtained by applying stationary models. In addition, a modeling result obtained by applying the $R2NGINAR(1)$ model is also placed in this table. Further, Table \ref{tabela221} contains results obtained by applying $R2NGINAR_{max}(p)$ and $R2NGINAR_{1}(p)$ models of various orders. Based on the tables, the following conclusions can be drawn. The weakest results are obtained by applying stationary models. Involving the concept of random environment  with $2$ different environment states into the modeling procedure brings significant improvement. This confirms the hypothesis that the time series really took place in two different environment states. Taking into account results given in Table \ref{tabela222}, $R2NGINAR(\mathcal{M,A,P})$ models are the best for selected real-life data among all models in random environment. In other words, selected real-life data may be observed as a realization of the generalized random environment $INAR$ time series of higher order. Therefore, it makes sense to test the effectiveness of the new \envst\ method on selected real-life data.

\begin{table}[htbp]
\caption{$CML$ parameter estimates and $RMS$-s obtained after application of various models on selected real-life data }{%
\begin{center}{\footnotesize\begin{tabular}{||l|l|l||l|l|l||}\hline\hline
Model&$CML$&$RMS$&Model&$CML$&$RMS$\\ \hline
\rule{0pt}{4ex}
$PoINAR(1)$&$\widehat\lambda=2.062$& 6.904&$GPQINAR(1)$&$\widehat\lambda=0.422$&7.097\\
&$\widehat\alpha=0309$& & &$\widehat\theta=0.825$&\\
& & & &$\widehat\rho=0.195$&\\
\hline\hline
\rule{0pt}{4ex}
$GINAR(1)$&$\widehat q=0.829$&7.028&$NGINAR(1)$&$\widehat\mu=4.573$&6.923\\
&$\widehat\alpha=0.286$& & &$\widehat\alpha=0.367$&\\
\hline\hline
\rule{0pt}{4ex}
$NGINAR(2)$&$\widehat\mu=4.573$&8.638&$NGINAR(3)$&$\widehat\mu=4.573$&8.679\\
&$\widehat\alpha=0.012$& & &$\widehat\alpha=0.011$&\\
&$\widehat p=0.184$& & &$\widehat p=0.184$&\\
\hline
\hline
\rule{0pt}{4ex}
$NGINAR(4)$&$\widehat\mu=4.573$&8.681&$NGINAR(5)$&$\widehat\mu=4.573$&8.671\\
&$\widehat\alpha=0.011$& & &$\widehat\alpha=0.016$&\\
&$\widehat p=0.141$& & &$\widehat p=0.139$&\\
\hline\hline
\rule{0pt}{4ex}
$NBRCINAR(1)$&$\widehat p=0.154$& 7.261&$RrNGINAR(1)$& $\widehat{\mathcal{M}}=(1.844,10.947)$& 5.745 \\
&$\widehat\rho=0,493$& & & $\widehat\alpha=0.145$ & \\
&$\widehat n=0,513$& & & & \\
\hline\hline \end{tabular}}\end{center}}\label{tabRealniPod}
\end{table}

\begin{table}
\caption{$CML$ parameter estimates and $RMS$-s obtained after application of $R2NGINAR_{max}$ and $R2NGINAR_{1}$ models of various orders on selected real-life data }
\begin{center}\label{tabela221}{\footnotesize\begin{tabular}{|c|c|c|c|}
\hline\hline
  \multicolumn{2}{c}{ $R2NGINAR_{max}(2)$   }\vline &\multicolumn{2}{c}{ $R2NGINAR_1(2)$ }\\
\hline\hline
$CML$ & $RMS$ & $CML$ & $RMS$ \\
\hline
\rule{0pt}{4ex}
 $\widehat{\mathcal{M}}=(0.912,7.204)$ & 6.408 & $\widehat{\mathcal{M}}=(0.913,7.204)$ & 6.409\\
 $\widehat\alpha=0.107$ & & $\widehat\alpha=0.107$ & \\
  $\widehat{\phi}=\left[\begin{array}{cc}
1 & 0\\
0.427 & 0.573
\end{array}\right]$ & & $\widehat{\phi}=(0.427,0.573)$ & \\
  \hline\hline
\multicolumn{2}{c}{ $R2NGINAR_{max}(3)$   }\vline &\multicolumn{2}{c}{ $R2NGINAR_1(3)$ }\\
\hline\hline
$CML$ & $RMS$ & $CML$ & $RMS$ \\
\hline
\rule{0pt}{4ex}
 $\widehat{\mathcal{M}}=(0.723,30.149)$ & 4.206 &$\widehat{\mathcal{M}}=(0.833,29.112)$ & 4.204 \\
  $\widehat\alpha=0.007$ & & $\widehat\alpha=0.008$ & \\
 $\widehat{\phi}=\left[\begin{array}{ccc}
1 & 0 & 0\\
0.998 & 0.002 & 0 \\
0.247 & 0.367 & 0.386
\end{array}\right]$ & & $\widehat{\phi}=(0.352,0.439,0.209)$ & \\
  \hline\hline
\multicolumn{2}{c}{ $R2NGINAR_{max}(4)$   }\vline &\multicolumn{2}{c}{ $R2NGINAR_1(4)$ }\\
\hline\hline
$CML$ & $RMS$ & $CML$ & $RMS$ \\
\hline
\rule{0pt}{4ex}
 $\widehat{\mathcal{M}}=(0.702,30.180)$ & 4.212 &$\widehat{\mathcal{M}}=(0.710,30.008)$ & 4.201 \\
  $\widehat\alpha=0.009$ & & $\widehat\alpha=0.009$ & \\
 $\widehat{\phi}=\left[\begin{array}{cccc}
1 & 0 & 0 & 0\\
0.970 & 0.030 & 0 & 0 \\
0.332 & 0.389 & 0.279 & 0\\
0.331 & 0.235 & 0.433 & 0.001
\end{array}\right]$ & & $\widehat{\phi}=(0.265,0.450,0.131,0.154)$ & \\
  \hline\hline
\multicolumn{2}{c}{ $R2NGINAR_{max}(5)$   }\vline &\multicolumn{2}{c}{ $R2NGINAR_1(5)$ }\\
\hline\hline
$CML$ & $RMS$ & $CML$ & $RMS$ \\
\hline
\rule{0pt}{4ex}
 $\widehat{\mathcal{M}}=(0.662,30.181)$ & 4.173 &$\widehat{\mathcal{M}}=(0.669,30.179)$ & 4.169 \\
  $\widehat\alpha=0.007$ & & $\widehat\alpha=0.007$ & \\
 $\widehat{\phi}=\left[\begin{array}{ccccc}
1 & 0 & 0 & 0 & 0\\
0.822 & 0.178 & 0 & 0 & 0\\
0.332 & 0.392 & 0.276 & 0 & 0\\
0.239 & 0.194 & 0.335 & 0.232 & 0\\
0.001 & 0.246 & 0.160 & 0.346 & 0.247
\end{array}\right]$ & & $\widehat{\phi}=(0.258,0.432,0.130,0.140,0.040)$ & \\
  \hline\hline
\end{tabular}.}
\end{center}
\end{table}

\end{document}